\documentclass[twocolumn,showpacs,preprintnumbers,amsmath,amssymb]{revtex4}

\usepackage{pstricks}
\usepackage{graphicx}
\usepackage{dcolumn}
\usepackage{bm}

\begin{document}

\newcommand{\locsection}[1]{\setcounter{equation}{0}\section{#1}}
\renewcommand{\theequation}{\thesection.\arabic{equation}}

\def\F{{\bf F}}
\def\A{{\bf A}}
\def\J{{\bf J}}
\def\af{{\bf \alpha}}
\def\beqn{\begin{eqnarray}}
\def\eeqn{\end{eqnarray}}

\def\dspace{\baselineskip = .30in}
\def\beq{\begin{equation}}
\def\eeq{\end{equation}}
\def\bw{\begin{widetext}}
\def\ew{\end{widetext}}
\def\pl{\partial}
\def\na{\nabla}
\def\al{\alpha}
\def\bt{\beta}
\def\Ga{\Gamma}
\def\ga{\gamma}
\def\de{\delta}
\def\De{\Delta}
\def\da{\dagger}
\def\ka{\kappa}
\def\si{\sigma}
\def\Si{\Sigma}
\def\te{\theta}
\def\La{\Lambda}
\def\lam{\lambda}
\def\Om{\Omega}
\def\om{\omega}
\def\ep{\epsilon}
\def\non{\nonumber}
\def\sq{\sqrt}
\def\sqg{\sqrt{G}}
\def\sp{\supset}
\def\sb{\subset}
\def\l{\left (}
\def\r{\right )}
\def\lq{\left [}
\def\rq{\right ]}
\def\fr{\frac}
\def\la{\label}
\def\hs{\hspace}
\def\vs{\vspace}
\def\inf{\infty}
\def\ran{\rangle}
\def\lan{\langle}
\def\ov{\overline}
\def\tl{\tilde}
\def\tm{\times}
\def\lrar{\leftrightarrow}


\preprint{HD-THEP-08-08}

\preprint{OSU-HEP-08-01}


\vs{1cm}

\title{Generation Symmetry and $E_6$ Unification}

\author{Berthold Stech}
\email{b.stech@thPhys.uni-heidelberg.de}

\affiliation{Institut f\"ur Theoretische Physik, Philosophenweg 16, D-69120 Heidelberg, Germany}
\author{Zurab Tavartkiladze}%
 \email{zurab.tavartkiladze@okstate.edu}

\affiliation{Department of Physics, Oklahoma State University, Stillwater, OK 74078, USA}

\vs{1cm}

\date{February 6, 2008}

\begin{abstract}

The group $E_6$ for grand unification is combined with the generation symmetry group $ SO(3)_g$. The coupling matrices
in the Yukawa interaction are identified with the vacuum expectation values of scalar fields which are representations of the
generation symmetry. These values determine the hierarchy of the fermions as well as  their mixings and CP-violation. This generation mixing appears in conjunction with the mixing of the standard model fermions with the heavy fermions present in the lowest representation of $E_6$. A close connection between charged and neutral fermions is observed  relating for instance the CKM mixings with the mass splittings  of the light neutrinos.
Numerical fits with only few parameters reproduce quantitatively all known fermion properties. The model predicts an inverted neutrino hierarchy and gives rather strict values for the light and heavy neutrino masses as well as for the $0\nu 2\beta $ decay parameter. It also predicts that the masses of the two lightest of six `right handed' neutrinos lie in the low TeV region.

\end{abstract}

\maketitle

\section{Introduction}\label{sec:1}

The origin of the properties of quarks and leptons, the masses and the mixings of the three generations, forms still an open problem in particle
physics. Grand unified theories \cite{Pati:1974yy} provide a general understanding of the structure and the quantum numbers of the standard model states and suggest new ideas.
The hope is to find a consistent scheme which provides for intimate relations between the many observables.   Lagrangians considered for a single
generation have the fundamental property of chiral symmetry, i.e. they are invariant under chiral transformations of the fermion fields taken together
with appropriate transformations of the Higgs fields.  For  more generations chiral symmetry requires an extended symmetry which relates the different generations: the fermions should be members of a non-Abelian generation symmetry..
There are two possibilities to enforce generation symmetry: one can either enlarge the number of Higgs fields by giving them generation indices or one can identify the  coupling matrices  in front of the Higgs fields as vacuum expectation values of new scalar fields carrying generation quantum numbers.

We choose here the second alternative in connection with the grand unified symmetry group $ E_6 $  \cite{E6}-\cite{recentE6} . For three generations the coupling matrices are $3\tm 3$ matrices. The corresponding 9 scalar fields are taken to be
hermitian and are expressed in terms of the hermitian $3\tm 3$ matrix $\Phi(x)$
\begin{eqnarray}
\label{1.1}
\Phi(x) = \chi(x) + {\rm i} \xi(x)~.
\end{eqnarray}
Here $\chi$ denotes a symmetric and $\xi $ an antisymmetric $3\tm 3$  matrix.
Clearly, this choice implies that conventional Yukawa interactions containing these fields are effective ones with dimension 5 and thus have to be understood on a deeper level.
This can be done by the introduction of additional heavy spinor fields. We will see, that by integrating out these heavy fields one finds interesting consequences for the
relation between quark and neutrino mass matrices.

The introduction of the hermitian matrix field $\Phi(x)$ coupled to the fermion fields suggests to use the group $SO(3)_g$ to describe the generation symmetry. In addition we will make use of a discrete parity like symmetry, generation parity $P_g$.
From the point of view of chiral symmetry  the use of $SU(3)_g$ instead of $SU(2)_g \simeq SO(3)_g $ would be more consequent.
We will not treat this extension because the subgroup $SO(3)_g$ of $SU(3)_g$, in combination with $P_g$, is sufficient to reach our aims. Together with the unification group $E_6 $ it can be dealt with in a very economical way.
 In the literature non-Abelian  continuous \cite{nonAbelGenSym} and discrete \cite{A4} flavor symmetries have been discussed and applied in various models.
 In our approach ${\bf all} $ fermion fields
are taken to transform as 3-vectors under the generation group $ SO(3)_g $. Consequently, the symmetric part of $\Phi$  has to transform as ${\bf 1 + 5}$ and the antisymmetric part
as ${\bf 3}$. Spontaneous symmetry breaking of this generation symmetry leads to vacuum expectation values of $\Phi$. By an orthogonal
transformation the symmetric matrix $\lan \chi \ran $ can be taken diagonal. As we will see its elements describe the up quark hierarchy:
\begin{equation}
\label {G}
\fr{\langle \chi \rangle }{M}= \hs{-0.8mm}G= \hs{-0.8mm}\fr{1}{m_t}\hs{-0.8mm}\left(\begin{array}{lll}
m_u&0&0\\
0&m_c&0\\
0&0&m_t
\end{array}\right)\hs{-0.1cm}=\hs{-0.1cm}\left(\begin{array}{lll}
\sigma^4&0&0\\
0&\sigma^2&0\\
0&0&1
\end{array}\right)~.
\end{equation}
Here $M$ denotes the scale at which the appropriate effective Yukawa interaction of dimension 5 is formed.
We take the mass ratios $m_c/m_t = m_u/m_c = \sigma^2$  to be valid at a high scale. At the weak scale $ M_Z$  these ratios are  somewhat modified. Taking $\si = 0.050$ gives good agreement with the experimental mass determinations. In the following we will use this parameter for expressing small quantities.

There remains only the vacuum expectation value of the antisymmetric matrix  $i\xi$  to produce all mixing and CP violating properties of quarks and leptons. (The use of a purely antisymmetric and hermitian mixing matrix was suggested in \cite{antE6}). We take for $\xi$ the particular form which was abstracted from the analysis of the fermion spectrum and the CKM matrix in \cite{Stech:2003sb} using $E_6$ symmetry for grand unification. It has a particular symmetry: it changes sign by exchanging the second with the third generation:
\begin{equation}
\label{A}
\frac{1}{M'}\langle {\rm i}\xi\rangle =
A= {\rm i}~ \lambda_A ~\left(
\begin{array}{ccc}
0&\sigma&-\sigma\\
-\sigma&0&1/2\\
\sigma&-1/2&0
\end{array}\right)~.
\end{equation}
The relative factor between the (1,2) and (2,3) elements determines the particle mixings. Its value $2 \sigma $ gives a good description of the mixings of quarks and neutrinos as will be shown in this article. The coupling constant $\lambda_A$ can be incorporated into the mass scale $M'$ (except for renormalization group  considerations).

As already mentioned above, besides the $SO(3)_g$  symmetry we also introduce a parity  symmetry $P_g$,  `generation parity'. The (extended) standard model fermions are taken to be even under this symmetry while the Higgs fields directly coupled to these fermion fields as well as the fields $\chi$ and $\xi$ have negative generation parity.
To obtain the spontaneous symmetry breaking of  $SO(3)_g \times P_g$  we  use in the Lagrangian for the field $\Phi ~~ SO(3)_g \tm P_g $ invariant potentials up to 4th order in the fields $\chi$ and $\xi$. Adding also the $SO(3)_g\tm P_g $ invariant Coleman-Weinberg  potential a complete breaking of the generation symmetry can be achieved. Moreover, by selecting properly the coefficients of these potentials the numerical values of 
(\ref{G}) and (\ref{A}) give the absolute minimum of the total potential.

It is clearly a challenge to obtain the masses, mixings and CP properties of all fermions with only the two generation matrices $G$ and $A$ together with a few vacuum expectation values of the corresponding Higgs fields. Suited for this task is the grand unification symmetry $E_6$ \cite{Stech:2003sb}. Here we can postulate that particle mixings are caused by the mixing of the standard model particles with the heavy particles occurring in the lowest representation ${\bf 27}$ of $E_6$. The up quarks cannot mix since they have no heavy partners in this representation. Thus, it follows immediately that - apart from a constant factor - their mass spectrum is simply given by $\lan \chi \ran /M = G $ [as presented in (\ref{G})].

As we will see the breaking of generation symmetry and $E_6$ symmetry
gives a hold on the complete fermion spectrum which includes new heavy
fermions. In particular, we find a very
strong hierarchy for the heavy right handed neutrinos: their mass matrix is
proportional to the square of the up quark mass matrix.  By integrating out
these heavy neutrinos, their masses appear in the denominator compensating the square of the Dirac mass matrix in the nominator. This mechanism gives the light neutrino spectrum a less pronounced hierarchy than the one of the up quarks. A possible consequence is an inverted neutrino spectrum. Such an interesting situation is not obtained in more conventional treatments of grand unified theories. It will be seen that very few parameters are sufficient to describe all the known properties of charged and neutral fermions in a quantitative way.

The use of $E_6$ as a grand unified theory has many virtues \cite{E6}-\cite{recentE6}. The fermions are in the lowest representation of this group and an elegant cyclic symmetry connects quark fields, lepton fields and anti-quark fields. In particular, as shown in \cite{Stech:2003sb}, $E_6$ combines the mixings among fermion generations with the mixing of the standard model particles with heavy charged and neutral states.

Thus, our starting symmetry at the GUT scale is
\begin{eqnarray}
\label{1.2}
E_6 \times SO(3)_g \times P_g~.
\end{eqnarray}
In order to introduce notations and conventions we add at this place a short description of the ${\bf 27}$ representation of $E_6$.
In the $E_6$ grand unification model the fermions are contained in the
${\bf 27}$
representation of the group, i e. they are described by ${\bf 27}$
two component  (left handed) Weyl fields for each generation:
\beq
\psi_r^{\alpha }~,~~~~
\alpha =1,2,3  ~~  , ~~~~   r=  1,\cdots ,27 ~.
\la{aa}
\eeq
$r$ denotes the $E_6$ flavor index and $\al $ labels the generations.
These fields - even under $P_g$ - describe the fermions of the standard model plus additional quark
and anti-quark fields with the same charge as the down quarks and new heavy
charged and neutral leptons.
All fermions are in singlet and triplet $SU(3)$ representations of the
maximal subgroup of $E_6$
\beq
\la{SU3}
 SU(3)_L  \times  SU(3)_R  \times SU(3)_C\equiv G_{333}~,
\eeq
which plays an important role in our approach.
In terms of $G_{333}$ we have
\beq
{\bf 27}=Q_L(x)+L(x)+Q_R(x)~,
\la{27dec}
\eeq
where the quantum number assignments are:
\beqn
{\rm Quarks}:~~~Q_L(x)=(3, 1, \bar 3)~,
\non
\\
{\rm Leptons}:~~~L(x)=(\bar 3, 3, 1)~,
\non
\\
{\rm Anti\hs{-0.3mm}-\hs{-0.3mm}quarks}:~~~Q_R(x)=(1, \bar 3, 3)~.
\la{ab}
\eeqn
{}For each generation one has
\beqn
\begin{array}{cc}
 & {\begin{array}{cc}
 \hspace{1.8cm}&
\end{array}}\\ \vspace{2mm}
(Q_L)_i^a=\hs{-0.2cm}
\begin{array}{c}
 \\

\end{array}\!\!\!\! &{\left(\begin{array}{ccc}
\hs{-0.3cm}u^a
\\
\hs{-0.2cm}d^a
\\
\hs{-0.2cm}D^a
\end{array}\hs{-0.2cm}\right)\! }~,\hspace{1.5cm}
\end{array}
\hs{-0.1cm}
\begin{array}{ccc}
& {\begin{array}{ccc}
 & &
\end{array}}\\ \vspace{2mm}
~~~~~~~~L^i_k= \hs{-0.5cm}
\begin{array}{c}
  \\
\end{array} \hspace{-0.1cm}&{\left(\begin{array}{ccc}
L^1_1 & E^{-} & e^{-}
\\
E^{+} & L^2_2 & \nu
\\
e^{+} & \hat{\nu } & L^3_3
\end{array}\right)~,
}
\end{array}
\non
\\
(Q_R)^k_a=\l \hat{u}_a,~\hat{d}_a,~\hat{D}_a \r ~,\hs{1.3cm}
\label {ac}
\eeqn
where $i, k, a=1, 2, 3$.
In this description $SU(3)_L$ acts vertically (index $i$) and $SU(3)_R$
horizontally (index $k$) and  $a$ is a color index. It is seen, that in each generation there exist 12 new fields extending the standard model: colored quarks
$D, \hat{D}$, charged  leptons $E^+, E^-$ and 4 new neutral leptons $\hat{\nu }, L_3^3, L_1^1, L_2^2$,
which all must correspond to heavy, but not necessarily very heavy, particles.

The group $G_{333}$ can serve as an intermediate gauge symmetry below the $E_6$ breaking scale (the unification scale). It can be unbroken only at and above the
scale where the two electro-weak gauge couplings combine. In the non-supersymmetric $E_6$ model, which we adopt here, this point occurs at $M_I
\simeq 1.3\cdot 10^{13}$~GeV according to the extrapolation of the standard model couplings. Interestingly, this is just the scale relevant for the neutrino masses using the
see-saw mechanism. Above this scale the united electroweak couplings and the QCD coupling run at first separately until the electroweak coupling bends to meet the
QCD coupling at the $E_6$ unification point. This happens at the
scale $\approx 3\cdot 10^{17}$~GeV (see Fig. \ref{fig-concorde} and more details in Appendix \ref{coup-unif}).

Before $ E_6$ symmetry breaking equivalent forms of (\ref{ac}) can be obtained by applying left and right $SU(3)$ ${\cal U}$-spin rotations. We fix the basis by using
vacuum expectation values (VEV) of the lowest Higgs representation,  the ${\bf 27}$ of $E_6$. In this representation possible vacuum expectation values are restricted
to the 5 neutral members sitting in positions corresponding to the ones of the neutral leptons in (\ref {ac}). Below the scale $M_I$ only two light $SU(2)_L$  doublets are
assumed to be active. We incorporate them in two different   scalar ${\bf 27}$-plet  fields. Thus we introduce two scalar fields $H_{27}$ and $\tl{H}_{27}$
with the following transformation properties under $E_6\tm SO(3)_g\tm P_g$
\beq
H_{27}\sim (27, 1, -)~,~~~~~\tl{H}_{27}\sim (27, 1, +)~.
\la{HtlH-transf}
\eeq
Only $H_{27}$ can couple to the $\psi$ fields in the form $ \chi ~\psi^T H \psi $.
Thus, it is convenient to choose a basis in which the VEV  $\langle H_{27} \rangle ^i_k $  forms a diagonal matrix.
In the scalar potential of
the $E_6$ Lagrangian the bilinear terms with respect to both scalar $\bf {27}$-  plets can contain the singlet part  $\chi_{(1)} $ of the fields
$\chi $:
\begin{equation}
\label{bilin}
\mu^2H_{27}^{\dag }H_{27}+\tl{\mu }^2 \tl{H}_{27}^{\dag }\tl{H}_{27}+\mu' \chi_{(1)} \l H_{27}^{\dag }\tl{H}_{27}+H_{27}\tl{H}_{27}^{\dag }\r ~.
\end{equation}
After $\chi_{(1)}$ develops a VEV, and for $\mu' $ sufficiently small (i.e. $|\mu'\lan \chi_{(1)} \ran| \ll |\mu^2-\tl{\mu }^2|$), the mass eigenstates
$H^u$ and $H^d$ following from (\ref{bilin}) are linearly related to $ H_{27} $ and $ \tilde H_{27} $
\beqn
H_{27} = H^u+z H^d~,~~~~~{\rm and}~~~~~\tilde H_{27}= -z H^u+H^d~,
\non
\\
~~~~{\rm with}~~~~z= -\fr{\mu'\lan \chi_{(1)}\ran }{(\mu^2-\tl{\mu}^2)} ~,~~~ |z| \ll 1.\hs{1cm}
\label{H-eigen-st}
\eeqn
We take the light up type doublet to be in
 $H^u$: $((H^u)^1_1,(H^u)^1_2) $ and the light down type doublet $(( H^d)^2_1, (H^d)^2_2) $ to occur in $H^d$.
These Higgs doublets are important for generating the correct fermion mass pattern.
$\lan (H_{27})^1_1\ran \simeq \lan (H^u)^1_1\ran = e^1_1$ determines the scale of the up quarks and
$\lan (H_{27})^2_2\ran \simeq z\lan (H^d)^2_2 \ran = z \epsilon^2_2 $ the scale of the down quarks and charged leptons of the standard model.  $ \epsilon^2_2$   is expected to be of the same order of magnitude as $e^1_1$. The factor $z$, which vanishes before $P_g$-symmetry
breaking, is responsible for the  small values of bottom and tau masses compared to the top mass.
Components of $H^u$, $H^d$ which are standard model singlets have large VEVs. For instance $\lan (H_{27})^3_3 \ran \simeq \lan (H^u)^3_3\ran =e^3_3$
 provide high (Dirac type) masses for all new quarks and leptons with the exception of the
 `right handed' neutrinos  $\hat{\nu }= L^3_2$ and $L^3_3$. Their masses arise from a different mechanism involving both $H_{27}$ and $\tilde H_{27}$.
  $e^3_3$ is not  a new scale parameter but is determined by the onset of $G_{333}$, the meeting point of the electroweak gauge couplings: $g_1(M_I) = g_2(M_I) $
with the result (see Appendix \ref{coup-unif}):

 \beq
 \label {e33}
 M_I = 1.27\cdot 10^{13}~{\rm GeV}~,~~~~~~ e^3_3 =\fr{M_I}{g_2 (M_I)}= 2.27\cdot 10^{13}~{\rm GeV}~.
 \eeq

We will see that this value of $e^3_3$  provides for the correct mass scale of the light neutrinos.

Due to the linear Yukawa coupling all fermions - except the 'right handed' neutrinos $L^3_2$ and $L^3_3 $  -  are Dirac particles and (so far) have the
 same $G$ hierarchy as the up quarks. The hierarchy of the 'right handed' neutrinos, on the other hand, becomes  super strong due to the combined action of $H_{27}$ and $ \tilde H_{27}$. Particle mixings will occur by taking into account  the antisymmetric generation matrix $ A $ combined with an antisymmetric (in $ E_6 $ indices) Higgs field. These mixings modify the hierarchy of down quarks and charged and neutral leptons in agreement with the experimentally observed particle spectra.

\section{Generation Symmetry}\label{sec:2}

For the construction of the Yukawa interaction symmetric under $E_6\tm SO(3)_g \times P_g $ the existence of a set of new fields is necessary.
First of all, as mentioned in the introduction, we introduce
scalar fields represented by the $3\tm 3$ matrix $\Phi(x)$. The members of the symmetric part $\chi $ transforms as $(1, 1+5, -)$
[under  $E_6\tm SO(3)_g \times P_g $] and the ones of the antisymmetric part $\xi $ according to $(1,3, -)$.
To have a renormalizable interaction we need new heavy vector like fermionic fields $F, \bar F$ and $F', \bar F'$  which transform as
\beqn
\hs{-1.1cm}F\sim (27, 3, -)~,~~~~~\bar F\sim (\ov{27}, 3, -)~,
\non
\\
F'\sim (27, 3, -)~,~~~~\bar F'\sim (\ov{27}, 3, -)~.
\la{F-trans}
\eeqn	

The Yukawa interaction involving the Higgs fields arises from the vertices
\beq
\left(\psi^T(g_{(1)}\chi_{(1)}+g_{(5)}\chi_{(5)})\bar F\right)~,~~
M(F^T \bar F)~,~~\left( F^T H_{27} \psi\right)~.
\label{vertex1}
\eeq
In (\ref {vertex1}) we used matrix notation with regard to generation indices, but suppressed  $E_6$ indices and Clebsch Gordan coefficients. $\chi_{(1)}$ and $\chi_{(5)}$ denote the singlet and $5$-plet parts of $\chi$, respectively.  By integrating out the massive fields $F, \bar F$ one gets the wanted effective interaction
\begin{equation}
\label{H}
\frac1M\left(\psi^T(g_{(1)}\chi_{(1)}+g_{(5)}\chi_{(5)})
H_{27} \psi\right)~.
\end{equation}
The corresponding diagram is shown in Fig.\hs{0.2cm}\ref {fig-Fexch}a.~
Clearly, the antisymmetric matrix $\xi$ does not contribute to the symmetric (in $E_6$ indices) field $H_{27}$ . It couples, however to the  antisymmetric Higgs representation $H_A (351)$ of $E_6$ with negative generation parity. The vertices are
\beqn
\left( (\psi^T g_{(3)}i\xi\bar F'\big) ~, ~ M'(F'^T \bar F')~,~~ (F'^T H_{A351})\psi \right)
\non
\\
\hs{-1cm}\Rightarrow ~~\frac{g_{(3)}}{M'}\left(\psi^Ti\xi H_{A}\psi\right)~.\hs{1.3cm}~
\label{vertex2}
\eeqn
The corresponding diagram containing the heavy fields $F', \bar F'$   is shown in
Fig.\hs{0.2cm}\ref {fig-Fexch}b.

The $P_g$ symmetry forbids couplings such as $\psi H \psi $ , $F \tilde H \psi $, $FFH $ etc.
In the following we will not separate the singlet and 5-plet parts of $\chi$ but simply use the combination
$\chi$  as occurring in $\Phi$  (i.e $ g_{(1)} = 1/3 ~g_{(5)} $ ) and absorb $g_{(5)}$  and the coupling  $g_{(3)}$ in the values of $ M$ and $M^{\prime}$, respectively. Now
we can  introduce number valued generation matrices by taking the vacuum expectation values of $\chi$ and $\xi$ as described in the introduction:
\begin{equation}
\label{G,A}
\frac1M\langle \chi\rangle = G ~,~~~~~~~\frac{1}{M'}\langle {\rm i}\xi\rangle = A ~.
\end{equation}
As mentioned before, the generation matrix $G$ in Eq. (\ref{G,A}) combined with vacuum expectation values of the Higgs field
 $ H_{27} $   gives Dirac masses  to all fermions except the two heavy leptons $ L^3_2$ and $L^3_3$. The latter require Higgs fields transforming as $(6,\bar 6)$ with respect to $SU(3)_L \times SU(3)_R$, components, which are not contained in $H $ and $H_{A}$. Instead of introducing another high dimensional Higgs field it is plausible to use the Higgs
field $H_{27}$ together with  $ \tilde {H}_{27}$ .
This avoids the appearance of a new unknown generation matrix. With the help of the generation symmetry it
 is possible to derive the relevant generation matrix of the
'composite' $(6,\bar 6)$ plet in terms of the $G$ matrix.
 This increases the predictive power of the model.
  To obtain the corresponding effective interaction from a renormalizable interaction, another massive Dirac field is required.
It is a vector in generation space and an $E_6$ singlet: $N(1,3, +)$, $\tl{N}(1,3, -)$ with $N$ even under $P_g$
and $\tilde N $ odd under $P_g$. In addition we need a total singlet field $(1,1,-)$ which is odd under $P_g $. It can be identified with the singlet part $ \chi_{(1)} $ of $\chi$ with vacuum expectation value
 $\lan \chi_{(1)} \ran = M_N \simeq M $. The vertices are
\begin{equation}
\label {vertex3}
\left(F^T H_{27}^{\dagger} N\right)~,~~~\langle \chi_{(1)} \rangle (\tl{N} N) = M_N  (\tl{N} N)~,~~~ \left( \tl{N} \tilde H_{27}^{\dagger} F\right)~.
\end{equation}
Integrating out the fields $N, \tl{N}$ the vertex $(F^T H^{\dagger})(\tilde H^{\dagger} F)/M_N$ emerges.
Below the masses of the $F$ states  the effective Yukawa interaction
\begin{equation}
\label{H,H}
\frac{1}{M^2}\frac{1}{M_N} \left( \psi^T \chi H_{27}^{\dagger}\right)
 \left (\tilde H_{27}^{\dagger} \chi \psi \right)~
\end{equation}
is generated.
The corresponding diagram is shown in Fig.\hs{0.2cm}\ref{fig-Nexch}.
Because in  (\ref {H,H}) the $E_6$ indices are not shown, one has to keep in mind that the combinations $ \psi H^{\dagger} $ and $ \psi \tilde H^{\dagger}$
are $E_6$ singlets. It is easily seen that by this interaction only neutral leptons can get masses, notably the right handed neutrinos $L^3_2$ and $L^3_3$ .
They are coupled to the large elements of $H_{27}$ and $\tilde{H}_{27}$, which are standard model singlet fields and thus will get large vacuum expectation values. The
most interesting feature is however the appearance of the square of $\chi/M$. It implies that after generation symmetry breaking the mass hierarchy of these heavy
neutrinos is dramatic, namely equal to $G^2$, the square of the mass hierarchy of the up quarks which is already a very strong one.

 $H_{27}$ and $\tl{H}_{27}$, which both appear now in the Yukawa interaction,  can be expressed in term of
the mass eigenstates $H^u$ and $H^d$. From (\ref{H-eigen-st}) we have $H_{27}=H^u+zH^d$ and $\tl{H}_{27}=-zH^u+H^d$,
 with the mixing parameter $ |z| \ll 1$ .
We  choose the VEVs of $H^u$ and $H^d$ as follows
\beq
\label{vevHH}
\langle H^u \rangle \hs{-0.8mm}=\hs{-0.8mm}
\left(\begin{array}{lll}
e^1_1&~~0&~0\\
0&~~0&~0\\
0&\hs{-0.15cm}-z\epsilon^3_2&~e^3_3\end{array}
\right) ,~~
\lan H^d\ran \hs{-0.8mm}=\hs{-0.8mm}
\left(\begin{array}{lll}
0&0&0\\
0& \ep^2_2&0\\
0& \epsilon^3_2& 0
\end{array}
\right).
\end{equation}
Thus,  the VEV of the  Higgs field  $H_{27}$  has  the diagonal form taken by convention
 \beq
\label{vevHcomb}
\lan H_{27}\ran =\langle H^u+zH^d \rangle \simeq
\left(\begin{array}{lll}
e^1_1&~0&~0\\
0&z\ep^2_2&~0\\
0&~0&~e^3_3\end{array}
\right)~,
\end{equation}
while $\tl{H}_{27}$'s VEV has the structure
\beq
\label{vevtlHcomb}
\lan \tl{H}_{27}\ran \hs{-0.6mm}=\hs{-0.6mm}\lan H^d-zH^u\ran  \simeq
\left(\begin{array}{lll}
-ze^1_1&~~0&~~~~~0\\
~~~0& ~~\ep^2_2&~~~~~0\\
~~~0&~~ \ep^3_2&~ -ze^3_3
\end{array}
\right)~.
\end{equation}
Since $\lan H_{27} \ran $  has no off diagonal element one expects a large $(3,2)$ element for $\tl{H}_{27}$.
We take $\ep^3_2 \simeq M_N \simeq  M $, where $M$ is identified with $ e^3_3 \simeq M_I$. In principle, however, $\ep^3_2$ together with $M$ and $M_N$ could be of a lower scale  (but still $\gg M_Z $).
The dominant VEVs of $ H_{27}$ and  $\tilde H_{27}$ fix now the mass terms of the `right handed' neutrinos $L^3_2$ and $L^3_3$
$$
F^{\{ 2, 3\}} G^2 ((L^3_2)^T L^3_3) + F^{\{ 3, 3\}} G^2 ((L^3_3)^T L^3_3)~,
$$
\beq
{\rm with}~~ F^{\{ 2, 3\}}\simeq \frac{e^3_3 \ep^3_2}{M_N} \simeq e^3_3 ~,~~ F^{\{ 3, 3\}} \approx  \frac{-z (e^3_3)^2}{M_N} \approx - z e^3_3 ~.
\label{F23,F33}
\eeq
$F^{\{ 3, 3\}}$ is the only Majorana mass term occurring so far.

The effective Yukawa interaction below $(M, M_N, M' )$ now reads
\beqn
{\cal L}^{\rm eff}_Y~ = ~G_{\alpha \beta}
\left(\psi^{\al T} H_{27} \psi^{\beta} \right)+ A_{\al \beta}
\left(\psi^{\al T} H_{A_{351}}\psi^{\bt}\right)+
\non
\\
\frac{1}{M_N}
\left( G^2 \right)_{\al \bt}
\left(\psi^{\al T} H_{27}^{\dagger}\tilde H_{27}^{\dagger} \psi^{\bt} \right)~.\hs{0.9cm}~
\label{eff}
\eeqn
This effective Yukawa interaction together with the VEV configurations (\ref{vevHcomb}) and (\ref{vevtlHcomb}) contains all the necessary
information about the generation structure: Generation hierarchy
and generation mixing are now completely fixed. At this stage we do not need to specify the scales of $M,  M_N \simeq M, M^{\prime} \gg M_Z$.
However, these scales become important when we take renormalization effects into account.

Let us now discuss the breaking of the generation symmetry $SO(3)_g$. The Lagrangian for the field $\Phi$ is
\beq
\label{LPhi}
{\cal L}_\Phi =\frac12 \rm{Tr}
\left\{\left(\partial_\mu\Phi-e[B_\mu,\Phi]\right)
\left(\partial_\mu\Phi-e[B_\mu,\Phi]\right)\right\}
-V(\chi,\xi)~,~
\nonumber
\eeq
with $B_\mu =B^i_\mu t^i$, where $B^i_{\mu}$ denote the vector potential and  $t^i$ the 3 antisymmetric generators of the generation symmetry $SO(3)_g$. $V(\chi,\xi)$ stands for an
$SO(3)_g$ invariant potential. Gauge invariance allows to choose $\chi$, the symmetric part of $\Phi$, to be a diagonal matrix with 3 real elements.
This defines a direction in symmetry space for a possible spontaneous symmetry breaking.
Normalizing $\Phi$ for this basis one can write
\begin{equation}
\label{diagonal}
\Phi=\left(\begin{array}{ccc}
\chi_1&0&0\\
0&\chi_2&0\\
0&0&\chi_3
\end{array}\right)+
\frac{\rm i}{\sqrt{2}}
\left(\begin{array}{ccc}
0&\xi_3&-\xi_2\\
-\xi_3&0&\xi_1\\
\xi_2&-\xi_1&0
\end{array}\right)~.
\end{equation}
In this form the vertices we have shown above simplify drastically since the orthogonal transformation diagonalizing $\chi $ can be absorbed by the fields $ \psi$ , $ F$ and $F'$ .

The scalar potential $V(\chi,\xi)$ in (\ref{LPhi}) can easily be made to have a minimum for specific values of the three invariants: the trace of $\chi^2$,
the trace of the square of the traceless part of $\chi$ and the trace of $\xi^2 $. At this minimum we have then two relations for the 3 fields forming  $\chi$ and one relation for the other 3 fields forming  $\xi $.  There  still remains the freedom of such $ SO(3)_g$ transformations which do respect this minimum and keep $ \chi $ diagonal. This remaining symmetry is necessarily a discrete subgroup of $ SO(3)_g$.  It is the discrete symmetry group $A_4$. As a generation symmetry $A_4$ has been suggested in many publications \cite{A4}. In our approach this symmetry occurs naturally in connection with the starting symmetry $SO(3)_g\tm P_g$. It is the remaining symmetry after using appropriate  potentials invariant under
$SO(3)_g\tm P_g$ and the choice of a symmetry breaking direction.

Radiative corrections add to the potential $V(\chi,\xi)$ new $SO(3)_g$ invariant parts which are of logarithmic type. The Coleman-Weinberg potential \cite{Coleman:1973jx} is of this form. By including it the total potential can lead to a complete spontaneous symmetry breaking of $SO(3)_g$ which results in vacuum expectation values for $\chi $ and $\xi $ of the form (\ref{G}), (\ref{A}). We demonstrate this here by using
the potential
\beqn
V(\chi,\xi)=-\fr{M^2}{2}{\rm Tr}[\chi^2]-\fr{{M'}^2}{4}{\rm Tr}[(i \xi)^2] +c_1({\rm Tr}[\chi ])^4+
\non
\\
c_T ({\rm Tr}[\chi_T^2] )^2+c_3{\rm Tr}[\chi^3]~{\rm Tr}[ \chi ]+
c_{\xi} ~({\rm Tr}[(i \xi)^2])^2+\hs{0cm}
\non
\\
c_{\chi\xi 0}{\rm Tr}[\chi.(i \xi)^2] ~{\rm Tr}[\chi ]+c_{\chi\xi1}{\rm Tr}[\chi.i \xi.\chi.i \xi]+ \hs{0.6cm}
\non
\\
\frac{3 e^2}{64 \pi^2} \left [ M_{B1}^4  \ln \l d_1 \fr{M_{B1}^2}{M^2}\r +
M_{B2}^4 \ln  \l d_2 \fr{M_{B2}^2}{{M'}^2}\r \right.
\non
\\
\left.  +M_{B3}^4  \ln \l d_3 \fr{M_{B3}^2}{{M'}^2}\r \right ]~.\hs{1.5cm}
\label {potential}
\eeqn
Here $\chi_T $ denotes the traceless part of $\chi$ and $M_{Bi}$ stands for the vector boson masses as field dependent functions
(we take in (\ref {potential})~ $ e=1 $ for simplicity). The coefficients in (\ref{potential}) can be tuned such that the $SO(3)_g\tm P_g$ symmetry breaks spontaneously and produces $G$ and $A$ with $\sigma = 0.050 $. To achieve this one has to use the six relations following from the first derivatives of the potential at the proposed minimum: $\lan \chi \ran =M G $ and
$ \lan \xi \ran = M^{\prime} A $.  Because of the large hierarchy a high accuracy of this  calculations is necessary. We use here $M^{\prime}/M=10^3$  i.e., the mass scale for forming the antisymmetric matrix $A$ is large compared to the one for forming the symmetric matrix $G$
(this is required for the gauge coupling unification, see the  renormalization group treatment in the appendix). Setting
then e.g. $ d_1=1$, $ d_2=237854/10^5$  and $ d_3= 237850/10^5 $  all other coefficients are determined by putting the first derivatives of the potential to zero.  These coefficients are sufficiently small  ($ c_1, c_T,  c_3, c_{\xi}  <1 $ for instance) to allow  perturbative treatments. The log terms are small near the minimum. With this choice of the potential the six eigenvalues of the $6 \tm 6$ matrix for the second derivatives of the potential are all positive at the wanted values of $\langle \Phi \rangle$. The minimum obtained is an absolute one (but degenerate with respect to different signs of the three elements in $\lan \xi \ran $).

The above formulae for the vacuum expectation values and the corresponding potential
 would have a different form if we would have used other values for the couplings $g_{(1)} $ and $g_{(5)} $.
 However, also for this more general situation  potentials can be constructed to produce  the required spontaneous symmetry breaking.

For the three vector boson masses at the minimum of the potential one gets
\begin{equation}
\label{vbmasses}
M_B \sim
\left\{\begin{array}{c}
 1.40~e M\\
0.505~e M^{\prime} \\
 0.505~e M^{\prime}
\end{array}\right\}~.
\end{equation}

\section{Charged Fermion Masses and Mixings}\label{sec:3}

As we have already mentioned, the diagonal generation matrix $\lan \chi \ran $ (i.e. $G$)
is taken such that the up quark masses have their observed hierarchy.
The top quark mass at $ M_Z$ is determined  by the vacuum expectation value $e^1_1$ and the coupling $\lambda_t$ at this scale
\beq
\label{tquark}
m_t = e^1_1 \lambda_t = e^1_1~G^{u\hat{u}}_{3,3} = 170.9~{\rm GeV}~.
\eeq
The notation (the superscript $u\hat{u}$ of $G$) indicates that at low energy one has to distinguish between different channels. The renormalization group effects are treated in the Appendix.
We take for $\lambda_t(M_Z) $ a value such that  the top coupling constant near $M_I$ is close to one.
\beq
\label{e11}
     \lambda_t(M_Z) = 1.30~~ , ~~~ e^1_1 = 131.5 ~{\rm GeV}~.
\eeq
Having for all diagonal $G$-couplings the same spacing at the scale $M_I$, we are able to calculate
the spacing for each channel at the scale $M_Z$. As obtained in the Appendix we find
\beqn
G^{u\hat{u}}(M_Z)={\rm Diag}\l r_t^{-1}\si^4 ~, r_t^{-1}\si^2~, 1\r \cdot \lam_t(M_Z)~,
\non
\\
{\rm with}~~~~r_t=0.717~.\hs{1.8cm}~
\la{GuuSpacing-at-MZ}
\eeqn
Comparing with the measured values of up and charmed quark masses suggest
\beq
\si =0.050~.
\la{sigma-value}
\eeq
Eq. (\ref{GuuSpacing-at-MZ}) gives for this value of $\sigma $ ~$m_u=1.5$~MeV and $m_c=0.60$~GeV in good agreement with experiment. We will use (\ref{sigma-value})
in all further calculations.

The case for the down quark and charged lepton masses is not as straightforward. Here  the light fermions will mix with the heavy states $D$ and $L$. We take this as the source of the particle mixings. It involves the indices $2$ and $3$ (${\cal U}$-spin mixing) of $SU(3)_L$  and $SU(3)_R$  of the Higgs field $H_A$. According to Eq. (\ref{eff}) this mixing of light and heavy fermions  occurs together with the generation mixings by the coupling matrix  $i \lan \xi \ran$ (i.e. the matrix $A$) .

\vs{0.2cm}

{\bf i) The down quark masses and mixings}

\vs{0.2cm}

The matrix elements for down quarks are part of a $6\tm 6$
mass matrix. It has the form

\begin{equation}
\label{MdD}
\begin{array}{cc}
 & {\begin{array}{cc}
 \hspace{0.5cm}\hat{d}\hspace{0.3cm}& \hspace{1.2cm}\hat{D}
\end{array}}\\ \vspace{2mm}
M_{d, D}=
\begin{array}{c}
 d\\D
\end{array}\!\!\!\!\!\! &{\left(\begin{array}{ccc}
\hs{-0.1cm}z\epsilon^2_2G+f_2^2 A~,&\hs{0.1cm} f_3^2 A
\\
\hs{-0.2cm}f_2^3A ~, &\hs{0.1cm}e^3_3G
\end{array} \right)\! }~.
\end{array}
\eeq
In this equation the constants $f^i_j$ stand for the
vacuum expectation values of $H_A(\bar 3, 3, 1)$:
\beq
 f^i_j = \lan(H_A)^i_j\ran~,~~~~ i, j=2, 3~.
\label{AVEV}
\eeq
$\epsilon^2_2$ and $f^2_3$ have  values of the order of the
weak scale, while $e_3^3$ and $f^3_2$ will be very large because they arise from standard model singlets. Further below we will find that  the vacuum expectation values of the standard model singlet components of $H_A$, like $f^3_2$, have values which are small compared to $e^3_3$ (smaller than  $\si^3 e^3_3$). This allows to make use of the see-saw formula and also justifies the neglection of $f^3_3 $ in (\ref{MdD}).
Integration of the heavy $D$-states leads to the $3\tm 3$ down quark mass matrix
at the scale $M_Z$
\beq
\label{renMd}
\hat{m}_d(M_Z)=z\epsilon_2^2G^{d\hat{d}}+A^{d\hat{d}} f_2^2 -
\fr{f_2^3f_3^2}{e_3^3}
A^{d\hat{D}}(G^{D\hat{D}})^{-1}A^{D\hat{d}}~,
\eeq
We distinguished different $G$ and $A$ matrices when coupled to different channels
$(d\hat d, d \hat D )$ etc.
Different renormalization factors arise when starting from
the original $G$
and $A$ matrices at high scales.  The calculations are deferred to the   appendices. The matrices appearing
in (\ref{renMd}) are defined in (\ref{GMZ}) and (\ref{AMZ}).

For the $d \hat d$ matrix element in (\ref{renMd}) with the generation index
$(3,3)$ we write
\beq
z \epsilon_2^2G^{d\hat{d}}_{33}= \epsilon_2^2\lam_b= m_b^0~.
\la{e22VEV}
\eeq
The down quark mass matrix leads to 7 observables :  3 mass eigenvalues, 3 mixing angles and the CP violating phase. According to (\ref{renMd}) we have 3 parameters for a fit of the experimental results. We use the notation of the various coupling constants provided in (\ref{GMZ}) and (\ref{AMZ}). Taking then
\beqn
m_b^0=2.95~{\rm GeV}~,~~~~~~
\lam_A^{d\hat{d}} f_2^2 \simeq  - 0.23~{\rm GeV}~,
\non
\\
\fr{f^3_2f^2_3}{e^3_3}
\fr{\lam_A^{d\hat{D}}\lam_A^{D\hat{d}}}{\lam^{D\hat{D}}}=
1.62\cdot 10^{-4}~{\rm GeV}~,\hs{0.4cm}~
\label{taking}
\eeqn
one gets
\beqn
m_d(M_Z)\simeq 2.6~{\rm MeV}~,~~~m_s(M_Z)\simeq 50~{\rm MeV}~,\hs{0.2cm}~
\non
\\
m_b(M_Z)\simeq 2.89~{\rm GeV}~,\hs{1.8cm}~
\non
\\
|V_{us}|\simeq 0.228~,~|V_{cb}|\simeq 0.042~,~|V_{ub}|\simeq 0.0039~.\hs{0.4cm}~
\la{downCKM}
\eeqn
The angles in the quark unitarity triangle have the values
\beq
\al_q \simeq 97^o~,~~~~~\bt_q \simeq 23^o~,~~~~~\ga_q \simeq 60^o~.
\la{Q-unit-trian}
\eeq
Our results are very satisfying. Quark masses, the CKM mixing angles and the three angles of the unitarity triangle  have values within experimental errors
\cite{Yao:2006px, Xing:2007fb}.

\vs{0.2cm}

{\bf ii) The charged lepton masses and mixings}

\vs{0.2cm}

The charged lepton sector is constructed in a similar way. The
light leptons mix with the heavy $L$'s through the vacuum expectation values
of the $ H_A(\bar 3, \bar 6, 1) $ multiplet (the vacuum expectation values of $ H_A( 6, 3, 1) $ are considered to be negligible because of its ${\bf 6}$ representation for the $SU(3)_L$ symmetry):
\beq
\lan H_A^{i\{j, k\}}\ran =f^{i\{j, k\}}~.
\la{A6VEVs}
\end{equation}
With the abbreviations
\beq
f^{2\{1,3\}}=g^2_2~,~~~
f^{2\{1,2\}}=g^2_3~,~~~f^{3\{1,3\}}=g^3_2
\la{abrev}
\eeq
and applying again the effective Lagrangian (\ref{eff})
the $6\tm 6$ mass matrix for charged leptons has the form
\begin{equation}
\begin{array}{cc}
 & {\begin{array}{cc}
 \hspace{0.5cm}e^{+}\hspace{0.9cm}& \hspace{0.9cm}E^{+}
\end{array}}\\ \vspace{2mm}
M_{e, E}=
\begin{array}{c}
 e^{-}\\E^{-}
\end{array}\!\!\!\!&{\left(\begin{array}{ccc}
-z \epsilon^2_2G-g^2_2A~,& ~~ g^3_2A
\\
-g^2_3A~, &~~-e^3_3G
\end{array} \right)\! }~,
\end{array}
\label{MeE}
\end{equation}
Integrating out the heavy $L$ states leads at the scale $\mu \simeq M_Z$ to the $3\tm 3 $ matrix
\beq
\hat{m}_e \simeq \hs{-0.4mm}-z\epsilon_2^2G^{e^{-}e^+}\hs{-0.4mm}-
g^2_2 A^{e^{-}e^{+}}\hs{-0.4mm} -\fr{g^2_3g^3_2}{e^3_3}
A^{e^{-}E^{+}}(G^{L\bar L})^{-1}A^{E^{-}e^{+}}.
\label{Me}
\eeq
Here appear again renormalization group coefficients  defined in (\ref{GMZ}), (\ref{AMZ}).
We choose
\beqn
z G_{33}^{e^{-}e^+} \epsilon^2_2 \simeq \epsilon ^2_2 \lam_{\tau }=m_{\tau }^0=1.593~{\rm GeV}~,
\non
\\
\lam_A^{e^-e^+}g^2_2 \simeq - 0.1898~{\rm GeV}~,\hs{0.6cm}~
\non
\\
\fr{g^2_3g^3_2}{e^3_3}
\fr{\lam_A^{e^{-}E^{+}}\lam_A^{E^{-}e^{+}}}{\lam^{L\bar L}}
\simeq 3.16\cdot 10^{-4}~{\rm GeV}~.
\la{inpLep}
\eeqn
These three numbers determine the charged lepton masses and, as in the quark case, determine also  the 3 mixing angles and the CP violating phase. The charged lepton mixings are not directly observable, but will play an important role in the discussion of the neutrino properties.
The masses come out precisely
\beqn
m_e(M_Z)=0.487~{\rm MeV}~,~~~~m_{\mu }(M_Z)=102.8~{\rm MeV}~,
\non
\\
m_{\tau }(M_Z)=1.747~{\rm GeV}~.\hs{2cm}~
\la{mefit}
\eeqn
For the mixing angles and the CP violating phase $\gamma^e $ we obtain
\beq
\te_{12}^e\simeq 4.8^o~,~~~~~\te_{23}^e\simeq 5.1^o~,~~~~~\te_{13}^e\simeq 0.35^o~,~~~~~\gamma^e\simeq -23^o~.
\la{ch-angles}
\eeq
We have defined these four parameters in complete analogy to the quark CKM mixing angles  and the unitarity angle gamma.

\vs{0.2cm}

{\bf iii) Estimates of $\epsilon^2_2$, $f^2_3$, $f^3_2$,
 $g^2_3$ and $g^3_2$

}

\vs{0.2cm}

So far only the combinations $ \lambda_{\tau} \epsilon^2_2$, $f^2_3 f^3_2$ and $g^2_3 g^3_2 $ are known numerically. However, there is one relation due to the known mass of the vector boson $W$ of the standard model. It connects all vacuum expectation values which belong to $SU(2)_L$ doublets:
\beq
\label{v}
(e^1_1)^2+(\epsilon^2_2)^2 + (f^2_2)^2+(g^2_2)^2+(f^2_3)^2+(g^2_3)^2\hs{-0.6mm}=\hs{-0.6mm}(174.1~{\rm GeV})^2~.
\eeq

Because of the nearly identical interaction of down quarks and charged leptons we can estimate the ratio
$f^2_3 / g^2_3 $ from the fit values given in (\ref {taking}) and (\ref {inpLep}) by taking the large standard model singlet VEV's $f^3_2$ and $g^3_2$ to be equal. One gets $\fr{f^2_3}{g^2_3} \simeq 0.5$ which is close to
$m_{\tau}^0 / m_b^0 $ as one could have expected. Using Eq. (\ref{v}) the only parameter needed then is the ratio  $ g^2_3 / \epsilon^2_2 $ . This is a ratio of two weak scale quantities with the same weak isospin quantum numbers. We therefore take it to be equal to one. This choice gives a value for $g^3_2 / e^3_3 $ small enough to justify the application of the see-saw mechanism we used above. It also provides for small values for the couplings $ \lambda_{\tau}$ and $\lambda_b$ necessary for renormalization group stability of the neutrino sector to be discussed in the appendices. All couplings and VEV's introduced are now fixed:
\beqn
f^2_3 \simeq 39~{\rm GeV}~,~~~~~g^2_3 =\ep^2_2\simeq 76~{\rm GeV}~,\hs{1cm}~
\non
\\
z G^{e^- e^+}_{3,3} \simeq \lambda_{\tau} \simeq\fr{ 1.6~{\rm GeV}}{\epsilon^2_2} \simeq 2\cdot 10^{-2}~, ~~\lambda_b \simeq 4\cdot 10^{-2}~,
\non
\\
\fr{f^3_2}{e^3_3} = \fr{g^3_2}{e^3_3} \simeq 4.16\cdot 10^{-6}~.\hs{1.8cm}~
\label{remainpara}
\eeqn
It is useful to set
$$ \fr{f^3_2}{e^3_3} = \fr{g^3_2}{e^3_3}= \sigma^3~x_g $$
giving
\beq
\label {x}
x_g \simeq 0.033 ~.
\eeq
The factor $\sigma^3 $ removes the singularities in the mass matrices with respect to the formal limit $\si \to 0$ .

\section{Neutrino Masses and Mixings}\label{sec:neutrino}

In each generation one has to deal with $5$ neutral leptons (see (\ref{ac})). Thus, the matrix for neutral leptons is a $15\tm 15$ matrix.
\begin{equation}
\begin{array}{ccccc}
 & {\begin{array}{ccccc}
\hs{0.3cm} L^2_3\hs{1.2cm} & \hs{0.7cm}L_2^3 \hs{1.2cm}
& \hs{0.6cm}L_3^3\hs{1.2cm} &\hs{0.6cm}L^1_1\hs{1.2cm} &
\hspace{0.6cm}L^2_2\hspace{0.2cm}
\end{array}}\\ \vspace{1mm}
M_L\hs{-0.1cm}=\hs{-0.1cm}
\begin{array}{c}
L^2_3 \\ L^3_2 \\ L^3_3 \\ L_1^1 \\ L^2_2
 \end{array}\!\!\!\!\! \hs{-0.1cm}&{\left(\begin{array}{ccccc}
~0~ & -e^1_1G~ &~ 0~ &~ -g^3_2A~ & ~0~
\\
-e^1_1G~   &~ 0 ~ &~ M_1~ &~ 0~ &~ 0~
 \\
~0~ &~M_1^T~ &~ M_2~ &~ 0~ &~ e^1_1G~
\\
-g^3_2A^T~ &~ 0~ &~ 0~ &~ 0~  &~ M_0~
\\
~0~ &~ 0~ &~ e^1_1G~ &~ M_0^T~ & ~ 0~
\hs{-0.1cm}\end{array}\hs{-0.1cm}\right)}
\end{array}  \!\!
\label{neu-ML}~.
\end{equation}
Each entry stands for a $3\tm 3$ matrix.  In the $12\tm 12$ sub-matrix for the heavy leptons we neglected small terms like $g^2_2$ and $g^2_3$. They play no role in the evaluation of the light neutrino properties. The following abbreviations are used:
\beq
\label{neu-M-entries}
M_0=e^3_3G~,~~~~M_1=F^{\{ 2, 3\}}G^2+F_A~A ~,~~~~M_2=F^{\{ 3, 3\}}G^2~.
\eeq
$ M_1$ and $M_2$ have the superstrong hierarchy $G^2$ according to ${\cal L}^{\rm eff}_Y$ and
Eq. (\ref{F23,F33}).
The only new element we had to introduce arises from the vacuum expectation value of $H_A( 6, 3, 1)$  which we argued to be negligible in the previous section. But here it appears together with very small elements occurring in
$ G^2 $ . Thus, the constant $F_A$ defined according to
$$ F_A =\langle (H_A)_{\{3,3\},1}\rangle = f_{\{3,3\}1}$$
is expected to be tiny compared to $F^{\{2,3\}}$ even though it is a singlet with respect to standard model transformations.
Nevertheless it has to be kept as a parameter.

It is useful to rewrite the matrix (\ref{neu-ML}) in the form
\begin{equation}
\begin{array}{ccc}
 & {\begin{array}{ccc}
\hspace{0.2cm} \hspace{0.7cm} &
&
\end{array}}\\ \vspace{1mm}
{M_L}=\hs{-0.2cm}
\begin{array}{c}
 \\  \\
 \end{array}\!\!\!\!\!\hs{-0.2cm} &{\left(\begin{array}{cc}
0 &\Om
\\
\Om^T   &\hat{M}
\end{array}\right)}~,
\end{array}  \!\!  ~~~~~
\label{neu-ML-2x2}
\end{equation}
where $0$ stands for the $3\tm 3$ zero block matrix, while $\Om $ and $\hat{M}$ are $3\tm 12$ and $12\tm 12$ matrices
respectively.
They read
\beqn
\Om= \l -e^1_1G,~ 0, ~-g^3_2A, ~0\r ~,
\la{neu-Om}
\\
%
\begin{array}{ccc}
 & {\begin{array}{ccc}
\hspace{0.2cm} ~\hspace{0.7cm} & \hspace{0.7cm} \hspace{0.9cm}
& \hspace{0.8cm}\hspace{1.2cm}
\end{array}}\\ \vspace{1mm}
\hat{M}=\hs{-0.2cm}
\begin{array}{c}
 \\  \\
 \end{array}\!\!\!\!\!\hs{-0.2cm} &{\left(\begin{array}{cccc}
0 &  M_1&0  &0
\\
M_1^T  &M_2 & 0 & e^1_1G
 \\
0 & 0 & 0 & M_0
\\
0 & e^1_1G & M_0^T & 0
\end{array}\right)}~.
\end{array}  \!\!
\label{neu-M}
\eeqn
The matrix $\hat{M}$ contains the masses of the heavy neutral states which should be integrated out. This can be done analytically since $g^3_2/e^3_3$ is sufficiently small according to (\ref{remainpara}).
In doing so the light neutrino $3\tm 3$ mass matrix is given by
\beq
m_{\nu }=-{\cal U}^T\Om \hat{M}^{-1}\Om^T{\cal U}~.
\la{neu-m-nu}
\eeq
In this expression the matrix $~{\cal U}~$ provides for a first order correction to the generalized see-saw result
$\Omega \hat{M}^{-1} \Omega^T$. One finds with sufficient accuracy
\beq
{\cal U} \simeq 1-\fr{1}{2}\Om^* \hat{M}^{-2}\Om^T~.
\la{neu-calU}
\eeq
The inverse of the matrix $\hat{M}$ is
\begin{equation}
\begin{array}{ccc}
 & {\begin{array}{ccc}
\hspace{0.2cm} ~\hspace{0.7cm} & \hspace{0.7cm} \hspace{0.9cm}
& \hspace{0.8cm}\hspace{1.2cm}
\end{array}}\\ \vspace{1mm}
\hat{M}^{-1}=\hs{-0.2cm}
\begin{array}{c}
 \\  \\
 \end{array}\!\!\!\!\!\hs{-0.2cm} &{\left(\begin{array}{cccc}
-\fr{1}{M_1^T}M_2\fr{1}{M_1}&  ~~\fr{1}{M_1^T}&~~-\fr{e^1_1}{M_1^T}G\fr{1}{M_0}  &~~0
\\
&&&
\\
\fr{1}{M_1} &~~0 &~~ 0 & ~~0
\\
&&&
 \\
-\fr{e^1_1}{M_0^T}G\fr{1}{M_1}  & ~~0 & ~~0 &~~ \fr{1}{M_0^T}
\\
&&&
\\
0 & ~~0 & ~~\fr{1}{M_0} & ~~0
\end{array}\right)}
\end{array}  \!\!  ~.
\label{neu-inv-M}
\end{equation}
From (\ref{neu-Om}) and (\ref{neu-inv-M}) one obtains
\beqn
\Om \hat{M}^{-1}\Om^T=
-(e^1_1)^2 \l G\fr{1}{M_1^T}M_2\fr{1}{M_1}G +\right. \hs{0.4cm}
\non
\\
\left.g^3_2G\fr{1}{M_1^T}G\fr{1}{M_0}A^T
+g^3_2A\fr{1}{M_0^T}G\fr{1}{M_1}G \r ~.
\la{OmMOm-gen}
\eeqn

Fig. \ref{fig-d5ops} shows how the terms in (\ref{OmMOm-gen}) are generated.
The generation of the first term in (\ref{OmMOm-gen}) is exhibited by the diagram of Fig. \ref{fig-d5ops}a, while the diagram of Fig. \ref{fig-d5ops}b
and its transpose shows the formation of the second and third terms of (\ref{OmMOm-gen}),
respectively.
In linear approximation with respect to $F_A$ the matrix $ M_1^{-1}$ becomes
\beq
M_1^{-1}\simeq \fr{1}{F^{\{ 2, 3\}}}\l \fr{1}{G^2}-\fr{F_A}{F^{\{ 2, 3\}}}\fr{1}{G^2}A\fr{1}{G^2} \r ~.
\la{invM1}
\eeq
The expression of ${\cal U}$ appearing in Eq. (\ref{neu-m-nu}) can be approximated by
\beq
\label{neu-calU-form}
{\cal U}\simeq 1-\fr{1}{2} g^3_2 A \fr{1}{M_0^T}\fr{1}{M_0} g^3_2 A \simeq \hs{-0.2cm}
{\left(\begin{array}{ccc}
1 &0 & 0
\\
0  & 1-\fr{x_g^2}{2}  & \fr{x_g^2}{2}
\\
0& \fr{x_g^2}{2}  & 1-\fr{x_g^2 }{2}
\end{array}\right)}~.
\eeq
For consistency the quantity $x_g $, at this place responsible for the correction to the see-saw formula, must be small compared to $1$ which is indeed the case as we have seen before.
Finally, taking all together, one obtains for the light neutrino mass matrix
\beqn
m_\nu= \fr{(e_1^1)^2}{F^{\{ 2, 3\}}}  {\cal U}^T   \l \fr{F^{\{ 3, 3\}}}{F^{\{ 2, 3\}}}{\bf 1}+\fr{g^3_2}{e^3_3}(A\fr{1}{G}+\fr{1}{G}A^T)-\right.
\non
\\
\left. \fr{F_A}{F^{\{ 2, 3\}}} \fr{g^3_2}{e^3_3}(A\fr{1}{G^2}A\fr{1}{G}+\fr{1}{G}A^T\fr{1}{G^2}A^T)\r {\cal U}~.\hs{0.5cm}
\label{mnu0}
\eeqn
In order to compensate the powers of $\si$ in the denominators of (\ref{mnu0})  we extract powers of $\sigma$ from $F_A$ as we did for $g^3_2$ .
\beq
\label{x_g,x_A}
\fr{g^3_2}{e^3_3}=\si^3 x_g ~~,~~~~ \fr{F_A}{F^{\{2,3\}}} = \si^5 x_A~.
\eeq
Like $x_g $ also $x_A$ needs to be small compared to one.

For a convenient description and discussion we use now the abbreviations
\beqn
m= x_g\fr{(e_1^1)^2}{F^{\{2,3\}}}~,~~~~~~~\rho =
\fr{F^{\{3,3\}}}{x_g F^{\{2,3\}}}~.\
\la{neu-mnu-params}
\eeqn

Let us at first discuss the case with only linear terms in $x_g$ and $\sigma$, putting also  $x_A=0 $ . The neutrino
mass matrix takes then the very simple form
\beq
\label{mnu1}
\begin{array}{ccc}
 & {\begin{array}{ccc}
 & &
\end{array}}\\ \vspace{1mm}
m_{\nu }\simeq \rho~m~{\bf 1} + \hs{-0.2cm}
\begin{array}{c}
 \\  \\
 \end{array}\!\!\!\!\!\hs{-0.2cm} & m~{\left(\begin{array}{ccc}
0 &-{\rm i} & {\rm i}
\\
-{\rm i}  &0   & -{\rm i}\fr{\si}{2}
 \\
 {\rm i}   & -{\rm i}\fr{\si}{2} & 0
\end{array}\right)}~.
\end{array}  \!\!  ~
\eeq

The eigenvalues of $m_{\nu}.m^{\dagger}_{\nu}$ are to first order in $\sigma $
\beqn
(m_2)^2 \simeq (\rho^2 +2 + \fr{\sigma}{\sqrt{2}})~ m^2~,
\non
\\
(m_1)^2 \simeq (\rho^2 +2 - \fr{\sigma}{\sqrt{2}})~ m^2~,
\non
\\
(m_3)^2 \simeq \rho^2 m^2~.\hs{1cm}~
\label{eival}
\eeqn
It is now easy to see the following properties of the light neutrinos:

{\bf i)}  The neutrino mass spectrum has the inverted form.

{\bf ii)} In the no mixing limit $x_g \to 0$  the 3 neutrino masses are degenerate.

{\bf iii)} The ratio between the solar and atmospheric mass squared differences is
$\fr{\si }{\sq{2}}\simeq  0.035$ independent of the parameters  in good agreement with experiments.

{\bf iv)} The experimentally observed atmospheric mass squared difference can be used to fix the mass parameter $ m$ for the light neutrinos and to estimate the mass parameter $F^{\{2,3\}}$ for the heavy neutrinos:
\beqn
 m \simeq \frac{1}{\sqrt{2}} \sqrt{\Delta m^2_{\rm atm}} \simeq 0.035~{\rm eV}~,\hs{1cm}
\non
\\
F^{\{2,3\}} \simeq \sqrt{2}x_g \fr{(e^1_1)^2} {\sqrt{\Delta m^2_{\rm atm}} }\approx 1.8\cdot 10^{13}~{\rm GeV}~.
\label{m,F}
\eeqn
Here we used for $x_g$ the estimate (\ref{x}). Thus, as expected from (\ref{F23,F33}), the largest heavy neutrino mass $F^{\{2,3\}}$ has indeed a value close to the intermediate mass scale $M_I$ !

{\bf v)} In the approximation used so far, and without taking renormalization effects into account, the neutrino mixing matrix is of the
{\it bimaximal} form.

In the next subsection we take renormalization into account by running the mass matrix down from $M_I$ to $M_Z$.
Moreover, the effect of $x_A$ and the correction terms $x_g^2$ (emerging from ${\cal U}$) will be included.

\vs{0.3cm}

{\bf Neutrino masses and mixings with RG effects}

\vs{0.2cm}

Now we make full use of Eq. (\ref{mnu0}) and take the renormalization effects  discussed in
Appendix \ref{ApB} into account.  Neglecting higher powers of the small quantity $\si $, the light neutrino
$3\tm 3$ mass matrix at the scale $M_Z$ has the form
\begin{widetext}
\beq
\label{mnuren}
\begin{array}{ccc}
 & {\begin{array}{ccc}
 & &
\end{array}}\\ \vspace{1mm}
m_{\nu }\simeq
 \begin{array}{c}
 \\  \\
 \end{array}\!\!\!\!\!\hs{-0.2cm} & m~{\cal U}^T~{\left(\begin{array}{ccc}
\rho (1+r_1) &-{\rm i} & {\rm i}
\\
-{\rm i}  &\rho (1+r_2)-2 x_A   & -{\rm i}\fr{\si}{2} r_{23} +x_A
 \\
 {\rm i}   & -{\rm i}\fr{\si}{2} r_{23} +x_A& \rho
\end{array}\right)}~\cal {U}~,
\end{array}  \!\!  ~
\eeq
\end{widetext}
where $r_1, r_2, r_{23}$ denote renormalization factors derived in  Appendix \ref{ApB} [see Eqs. (\ref{r12-RG-expr}), (\ref{r23-RG-expr})].
Their numerical values are
\beq
r_1=0.1325~,~~~~r_2=0.0582~,~~~~r_{23}=0.9426~.
\la{neu-RGr-factors}
\eeq

Without renormalization the value of $\rho$ has no influence on the neutrino mixing pattern. Now, however, $\rho $ plays a role. It should not be large  in order not to change the mixing pattern drastically.
In fact, $F^{\{3,3\}}/F^{\{2,3\}}$ must be as small as $x_g$ in accord with  $ \rho \approx -  z/ x_g $ (\ref{F23,F33}) and our value for  $x_g$~, e.i. $|\rho |$  must be around $1$. As a consequence, the inverted spectrum found above still prevails.

With  the following choice of parameters we obtain a very good description of the known neutrino properties.
In this fit we adjust $x_A$ and  $\rho$  and decrease $ m$  slightly. For  $ x_g$ we take the value used already.
\beqn
m=0.0324~ {\rm eV}~,~~~~~\rho=1.132~,
\non
\\
x_g = 0.033 ~,~~~~~x_A= -0.0601~.
\label{neu-par-ren}
\eeqn
For the neutrino masses one gets
\beqn
m_1=0.0616~{\rm eV}~,~~~~m_2=0.0623~{\rm eV}~,
\non
\\
m_3=0.0374~{\rm eV}~.\hs{1.3cm}~
\la{neu-mass-ren1}
\eeqn
Accordingly, the  mass square differences are
\beqn
\De m_{\rm sol}^2=m_2^2-m_1^2=8\cdot 10^{-5}~{\rm eV}^2~,
\non
\\
|\De m_{\rm atm}^2|=m_2^2-m_3^2=2.5\cdot 10^{-3}~{\rm eV}^2~.
\la{neu-mass-ren2}
\eeqn
For the mixing angles emerging from the neutrino mass matrix one finds
$\te_{12}^{\nu }\simeq 35.3^o, \te_{23}^{\nu }\simeq 39.4^o$ and ~$\te_{13}^{\nu }\simeq 3.2^o$.
By including the effect of the diagonalization of the charged lepton mass matrix  as obtained in section \ref{sec:3}, the neutrino mixing angles and the CP violating phase $\de_l$ (appearing in the neutrino oscillation amplitude) become
\beqn
\te_{12}\simeq 34^o~,~~~\te_{23}\simeq 43^o~,~~~\te_{13}\simeq 6.3^o~,~
\non
\\
~\de_l \simeq 67^o~.\hs{2.4cm}
\la{neu-angles-ren2}
\eeqn
The angles and the phase are given according to the standard parametrization.
These mixing angles together with the mass squared differences (\ref{neu-mass-ren2}) are in perfect agreement with
global fits to neutrino oscillation data \cite{Fogli:2005cq}. These results are very satisfactory.
For the neutrino less double $\beta $-decay parameter one finds
\beq
|\lan m_{\bt \bt }\ran |\simeq 0.046~{\rm eV}~.
\la{2btdec-ren}
\eeq
 We note that there is not much freedom within this $E_6$ model to get larger or smaller values for the neutrino masses or the  $0\nu 2\bt $-decay parameter,  or to change the neutrino hierarchy. We already observed  that $F^{\{2,3\}}$, the mass of the heavy third generation right handed neutrino turned out to be amazingly close to $M_I$,
the point where the electroweak gauge couplings of the standard model meet.

As a consequence, we know now, at least approximately, the masses and
mixings of all heavy fermions. A direct diagonalization of the $15\tm 15$
neutral lepton mass matrix  reveals in particular, that the two lightest
of the 'right handed' neutrinos (i.e.  the first generation)  turn out
to have masses of only $ \approx 700$~GeV. This is caused by the super
strong hierarchy valid for these particles. On the other hand, the
masses of the first generation of heavy quarks and heavy $SU(2)_L$ leptons
 are much larger ($\approx 10^8$~GeV). We also
find,  that the  neutrino mixing matrix for the light neutrinos is not
strictly unitary. The light neutrinos mix to about $2\%$ with the  neutral
leptons $L^1_1$ and  $L^2_2$. The two lightest of the heavy neutrinos (a combination of the first generations of $\hat{\nu }=L^3_2$ and $L^3_3$) mix
with an amplitude of about $8\cdot 10^{-7}$  with the first generation of the light
neutrinos. These heavy neutrinos could be detected by their decay to the Higgs  field $(H^u)^1_1$ [from the $SU(2)_L$ doublet
$\l (H^u)^1_1, (H^u)^1_2\r $] and a light neutrino provided the mass of the Higgs field $(H^u)^1_1$ lies below the heavy neutrino masses. If not, they can decay  into  $W$ and $Z$ gauge bosons  and a light lepton
via the mixings given above.

Our phenomenological treatment gives a clear picture of the spectrum and mixings of the standard model fermions and of  their heavy $E_6$ partners. A quantitative fit reproducing all known properties of the standard model  charged and neutral fermions could be performed. Several not yet measured quantities are predicted. However, the model is still incomplete, because an understanding of the scalar sector, in particular of the vacuum expectation values of the $E_6$ Higgs fields, is not yet achieved.

\section{Summary}\label{summary}

In this work we have addressed the problem of fermion masses and mixings. We took the gauge group  $E_6$ for grand unification augmented with the  generation symmetry $SO(3)_g\tm P_g$.
The fermion fields of the standard model are taken to be in the $(27,3,+)$ representation of $E_6 \tm SO(3)_g\tm P_g $.  The scalar fields transforming under the generation
symmetry obtain vacuum expectation values by  a complete spontaneous  symmetry breaking. The corresponding values provide for the hierarchy of the fermions, for their mixings and CP-violation. The generation mixings occur in conjunction with the $SU(3)$ ${\cal U}$-spin mixing of the standard model fermions with their heavy partners.
To have a renormalizable model we had to introduce heavy Dirac fields. Integrating them out led to a very strong hierarchy of the right handed heavy neutrinos with
important consequences for the light neutrino properties. The onset of the intermediate symmetry
$SU(3)_L \tm SU(3)_R \tm SU(3)_c$ occurs at the well-known meeting point of the electro-weak gauge couplings of the standard model. This scale determines the masses of the heavy neutrinos and turns out to be in full accord with the measured mass splittings of the light neutrinos. Also the formation of the Yukawa coupling matrices determining the fermion hierarchies can happen at this scale.  Our model needs only few fit parameters to reproduce all known masses and mixing properties of the fermions.
Because of the unique use of the $up$ quark hierarchy $G$ and the antisymmetric mixing matrix $A$ for quarks, charged leptons and neutrinos, the mixing in the
quark sector determines the masses, the splittings and the main part of the  mixings in the neutrino sector. We obtain for the light neutrinos the inverted hierarchy
and - for the unrenormalized case  - bimaximal mixing. Taking renormalization group effects and the mixings coming from the charged lepton sector into account and by adjusting the small parameter $x_A$,  the slight change
to the  experimentally observed mixing angles can be achieved. The entire spectrum of light and heavy fermions can be estimated.
The two light neutrinos with `solar mass splitting'
have masses  $\simeq 0.06$~eV, while the lightest one weights $\simeq 0.04$~eV. The angle $\te_{13} $ is predicted to be $\approx 6^o$.
The CP violating phase $\de_l$ appearing in neutrino oscillations  is large, around  $70^o $.
For the  parameter relevant for neutrino less double $\beta $-decays we find $\simeq 0.046 $~eV. The two lightest members of the heavy neutrinos have a surprisingly low mass of about $700$~GeV.  The neutrino mixing matrix is not fully unitary,  mixings to some high mass states have a magnitude of about 2 percent.

\vs{0.1cm}

\hs{-0.5cm}
{\bf Acknowledgments}
\vs{0.1cm}

\hs{-0.5cm}
B.S. likes to thank C. Hagedorn for a helpful discussion on discrete groups.
The  work of Z.T. is supported in part by the DOE grants DE-FG002-04ER41306 and DE-FG02-04ER46140.


\appendix

\renewcommand{\theequation}{A.\arabic{equation}}\setcounter{equation}{0}
\section{Gauge Coupling Unification and
Renormalization of Flavor Matrices}\label{appA}

\subsection{Gauge Coupling Unification}\label{coup-unif}

In the previous section we have fixed the field content of the fermion sector
which are non singlets under $E_6$ and contribute to the gauge coupling
running. Also,  scalar representations play an important role. In this
section we will give the full list of fields appearing at appropriate
energy scales. This allows to study gauge coupling unification and to examine
whether or not perturbativity is kept up to the
$M_{\rm Pl}\simeq 2.4\cdot 10^{18}$~GeV
(the reduced Planck mass).

We assumed that the scales $e^3_3$ and $\epsilon^3_2$ are both equal to $M_I/g_2(M_I)$  (at $M_I$
the couplings $g_1$ and $g_2$ meet). Below this scale the field
content consists of the three fermion generations of the standard model
together with two Higgs doublets. One light (up type) Higgs doublet $h_u=((H^u)^1_1, (H^u)^2_1)$ comes from
$H^u(\bar 3, 3, 1)$ and has the VEV $\lan (H^u)^1_1\ran =e^1_1$. The second one, the down type Higgs doublet
$h_d=((H^d)^1_2, (H^d)^2_2)$, emerges from $H^d(\bar 3, 3, 1)$ with the VEV $\lan (H^d) ^2_2\ran =\epsilon^2_2$.
Therefore, below the scale $M_I$,  the b-factors relevant for the gauge coupling constants
$g_1$, $g_2$, $g_3$  affected by these states are:
\beq
\l b_1, b_2, b_3\r =
\l \fr{21}{5}, -3, -7\r ~.
\la{bSM}
\eeq
In addition, there are extra $D$ and $L$ fermion states from the $27^{\al }$ fermion representation. Their masses have the scales
$ M_{I}(\si^4, \si^2, 1)$. The corresponding b-factors valid for each generation of $D$'s and $L$'s are
\beq
\hs{-1mm}\l b_1, b_2, b_3 \r^D\hs{-2mm}=\hs{-0.8mm}\l \fr{4}{15}, 0, \fr{2}{3}\r ,\hs{0.8mm}
\l b_1, b_2, b_3 \r^L\hs{-2mm}=\hs{-0.8mm}\l \fr{2}{5}, \fr{2}{3}, 0 \r .
\la{bAD}
\eeq
contributing at different scales.
Above $M_I$, all components of $27^{\al}$ can be considered light.
Together with these, there are `light' $H^u(\bar 3, 3, 1)$ and $H^d(\bar 3, 3, 1)$ scalars.
The b-factors corresponding to all these states are
\beq
\l b_{L}, b_{R}, b_C\r^{M_I} =\l -4, -4, -5\r ~.
\la{b333}
\eeq
Moreover, at the scale $M$ [the masses of the states
$F_{\al }(27), \ov{F}^{\al }(\bar 27)$ ] which we  identify here with $M_I$, the RG $b$-factors receive the additions
$b_i^F=(12, 12, 12)$.
Between the scales $M_I$ and $M_{\rm GUT}$  we have $L\leftrightarrow R$
(${\cal D}$-symmetry) which insures the equality $g_L(\mu )=g_R(\mu )$. It starts  at
 $M_I$, the meeting point of $g_1$ and $g_2$.
At some scale $M_6$ ($M_I<M_6<M_{\rm GUT}$)  the scalar states
$(6, 3, 1)$, $(\bar 3, \bar 6, 1)$ [arising from $H_A(351)$]  have to come in. We  take them with the common mass $M_6$.
The leptonic states from ${F_{\al}}', \bar{F^{\al }}'$
$(L_{F'}+\bar L_{\bar F'})_{\al }$ are also assumed to have masses below the GUT scale. For convenience
(see the discussion below) we will take their masses equal to $M_6$.
 All states with mass $M_6$
are important for the successful gauge coupling unification. The corresponding  b-factors are
\beq
\l b_{L}, b_{R}, b_C\r^6 =\l \fr{19}{2}, \fr{19}{2}, 0\r ~.
\la{b6}
\eeq
By taking $M_6=3.47\cdot  10^{16}$~GeV   unification is completed at
$M_{\rm GUT} =2.8\cdot 10^{17}$~GeV. The mass scale $M_6$ has been chosen in such a way as to insure the perturbativity
of the unified gauge coupling $\al_{\rm GUT}$  up to the Planck scale $M_{\rm Pl}$.
The plot which shows the unification of the gauge couplings
is given in Fig.\hs{0.1cm}\ref{fig-concorde}. It has a `Concorde' shape pretty similar to the one first obtained  in Ref. \cite{Stech:2003sb}.
For the case considered here we have
\beqn
M_I\simeq 1.27\cdot 10^{13}~{\rm GeV}~,~~M_6\simeq  3.47\cdot 10^{16}~{\rm GeV}~,~
\non
\\
M_{\rm GUT}\simeq 2.8\cdot 10^{17}~{\rm GeV}~,~\al_{\rm GUT}^{-1}(M_{\rm GUT})\simeq 24.1~.~~
\la{Unif-values}
\eeqn

At and above  $M_{\rm GUT}$ we include, together with the states mentioned above,  the remaining states of
${F_{\al }}'(27), {\ov{F}^{\al }}'(\bar 27)$ in order to complete the $E_6$ states. Together with all
$E_6$ gauge bosons, the complete $E_6$ scalar multiplets $H(27)$, $\tl{H}(27)$ and $H_A(351)$,   also the scalar field  $H(650)$ is taken into account.
The role of the $650$-multiplet is to break $E_6$ down to $G_{333}$ at the
scale $M_{\rm GUT}$. The b-factor above $M_{\rm GUT}$, corresponding to all these states,
is $b_{E_6}=63$.
With this input we can evolve  the $E_6$ unified gauge coupling above $M_{\rm GUT}$ and compute its value at the Planck mass:
\beq
\al_{\rm GUT}^{-1}(M_{\rm Pl})=\al_{\rm GUT}^{-1}(M_{\rm GUT})-\fr{63}{2\pi }\ln \fr{M_{\rm Pl}}{M_{\rm GUT}}~.
\la{alGUT-Pl}
\eeq
With the values of $M_{\rm GUT}$ and $\al_{\rm GUT}(M_{\rm GUT})$ given in (\ref{Unif-values}) we find from (\ref{alGUT-Pl})
 $\al_{\rm GUT}(M_{\rm Pl})/{4\pi }\simeq 0.03$ (which may be viewed
 as an effective loop expansion parameter).
It is seen, that within our approach the gauge couplings remain perturbative up to  the
Planck scale.

\subsection{Flavor Coupling Renormalization}\label{fl-ren}

Let us start with the renormalization of the top, bottom and tau Yukawa couplings. They are
generated at the scale ~ $ \lan \chi_{33}\ran \simeq M $ after integrating out the states
$F, \bar F$.
In order to have RG stability in the neutrino sector we need to have a rather small value for the
$\tau$ coupling constant $\lambda_{\tau}$.  This is satisfied because of the small value of $z$.
In a  notation appropriate for the $G_{333}$ symmetry, the Yukawa interaction involving the $G$ matrices reads
\beqn
G^{QQ}Q_LQ_RH^u(\bar 3,3,1)+zG^{QQ}Q_LQ_RH^d(\bar 3,3,1)+
\non
\\
\fr{1}{2}zG^{LL}LLH^d(\bar 3,3,1)~.\hs{1.3cm}~
\la{G-Yuk-G333}
\eeqn
Since the up type quark masses are generated by $e^1_1\subset H^u$ and the down type and charged lepton masses through
$\epsilon^2_2\subset H^d$, we have
\beq
\lam_t\hs{-0.5mm}=\hs{-0.5mm}(G^{QQ})_{33}~,~~\lam_b\hs{-0.5mm}=\hs{-0.5mm}z(G^{QQ})_{33}~,~~
\lam_{\tau }\hs{-0.5mm}=\hs{-0.5mm}z(G^{LL})_{33}~.
\la{tbtau-Yuk-GHtlH}
\eeq
We estimated $z G^{e\hat e}\simeq 0.02$ in (\ref{remainpara}) using $\epsilon^2_2 \simeq
76$~GeV. Thus, RG effects for $\lambda_{\tau}$ and $\lambda_b$ are small and the only strong scale dependence below $M_I$ occurs for $\lambda_t$ .
For a detailed description we will now use the  appropriate notation for all Yukawa couplings.
Namely, $e_1^1G\to e^1_1G^{u\hat{u}}$, $z\epsilon_2^2G\to z\epsilon_2^2(G^{d\hat{d}},  G^{e^-e^+})$,
$f_2^2A\to f_2^2A^{d\hat{d}}$ etc,
and  $(G^{u\hat{u}})_{33}=\lam_t$, $z(G^{d\hat{d}})_{33}=\lam_b$, $z(G^{e^-e^+})_{33}=\lam_{\tau }$,
$(A^{d\hat{d}})_{23}={\rm i}\lam_A^{d\hat{d}}/2$ etc.

In the charged lepton sector we had already fixed some Yukawa couplings and  products of Yukawa couplings with the corresponding VEVs.
The values for $\lambda_t$, $\lambda_b$ and $\lambda_{\tau}$ are
\beq
\lam_t(M_Z)=1.30~,~~~~ \lam_b(M_Z) \simeq 4\cdot 10^{-2}~,~~~~
\lam_{\tau }(M_Z) \simeq 2\cdot 10^{-2}~.
\la{Yuk-MZ}
\eeq

The Yukawa coupling in front of the matrices  $A$ is defined according to
\begin{equation}
\label{AlamA}
A={\rm i}\left(
\begin{array}{ccc}
0&\sigma&-\sigma\\
-\sigma&0&1/2\\
\sigma&-1/2&0
\end{array}\right)\lam_A ~.
\end{equation}
In our estimates we take for all $\lam_A$'s   the same value, however.

In our model in which the mixing of $E_6$ flavors is a ${\cal U}$-spin mixing effect, one can expect that
the VEV's of $\epsilon^2_2$, $f^2_3$ and $g^2_3$ which appear at the weak scale have roughly the same magnitude. This implies using (\ref{v}) and the discussion at the end of section \ref{sec:3} ~$\epsilon^2_2 \simeq g^2_3 \simeq 76$~GeV.
Together with the fit values  quoted in (\ref{taking}), (\ref{inpLep}) all VEV's  encountered are known at least approximately.
Using $\lam^{D\hat{D}}\approx \lam^{L\bar L} \approx 1$ and $\lambda_A=1$ we exhibit them here all together:
\beqn
e^1_1=131.5~{\rm GeV} ,~f^2_2\simeq -0.23~{\rm GeV} ,
~g^2_2\simeq - 0.19~{\rm GeV} ,
\non
\\
f^2_3\simeq 39~{\rm GeV}~,~~~~~g^2_3=\ep^2_2 \simeq 76~{\rm GeV}~,\hs{1cm}~
\non
\\
\hs{-1cm}\fr{f^3_2}{e_3^3} = \fr{g^3_2}{e^3_3} \simeq 4.16\cdot 10^{-6}~.\hs{1.5cm}~
\la{EW-VEVs}
\eeqn
It is important that the couplings $\lam_b$ and $\lam_{\tau }$ are small in this scenario. It implies that their RG dependence can be ignored. We could select $\lam_A \simeq 1$ because in the RG treatment at scales below $M_I$  this coupling
does not explicitly enter. The reason is, that the scalar states $(6,3,1)$ and  $(\bar 6,\bar 3,1)$   decouple already at the higher scale $M_6$ and thus cannot appear in appropriate loops.
The only renormalization effects relevant below the scale $M_I$  are due to $\lam_t$ and the gauge couplings
$g_{1,2,3}$. The renormalization of the flavor matrices is performed in the following paragraph  taking this fact into account.

The flavor matrices $G$ and $A$ as introduced in section \ref{sec:1} did not include
RG effects. We start with the study of the renormalization of the $G$ matrices which are generated at the scale $M_I$. At this point
the spacing of different channels (i.e. $u\hat{u}$, $d\hat{d}$, $e^-e^+$ etc)  is universal.  We suggested for the splitting equal spacing between first and second and between second and third generations  (\ref{G}).
 At the scale $M_Z$, however,  for different channels different RG factors  emerge
affecting the spacings. The
$G$-couplings can be divided into two groups.  One group corresponds to couplings involving $H^u_{27}$.
The second group corresponds to couplings involving $H^d_{27}$.
At $M_Z$ they are split as follows:~
$G_{H^u} \to \l G^{u\hat u} ,  G^{D\hat{D}}, G^{L\bar L}  \r $,
$zG_{H^d}\to z\l G^{d\hat{d}} , G^{e^-e^+},\cdots \r $ (we will keep only those quantities which are relevant for us).
Our aim is to calculate the relative factors between appropriate entries in these matrices. The relative RG factors between the 3rd generation components of $G^{u\hat{u}} ,  G^{d\hat{d}}$ and $G^{e^-e^+}$ are described  by the coupling constants $\lambda_t$, $\lambda_b$, $\lambda_{\tau}$.
However, different generations have different Yukawa interactions. Because we can  take $z\approx 0.02$  all Yukawa couplings involving this prefactor do not participating in the change of the spacings. With the boundary condition that all G-couplings at $M_I$ have the form (\ref{G}), one gets  at the scale $ M_Z$
\begin{eqnarray}
\fr{G^{u\hat{u}}}{\lam_t}(M_Z)={\rm Diag}\l r_t^{-1}\si^4~,~r_t^{-1}\si^2~,~1\r ~,
\nonumber
\\
\fr{G^{d\hat{d}}}{\lam_b}(M_Z)={\rm Diag}\l r_t^{-1/3}\si^4~,~r_t^{-1/3}\si^2~,~1\r ~,
\nonumber
\\
\fr{G^{e^-e^+}}{\lam_{\tau }}(M_Z)={\rm Diag}\l \si^4~,~\si^2~,~1\r ~,
\nonumber
\\
\fr{G^{D\hat{D}}}{\lam_D}(M_Z)={\rm Diag}\l \ka^D_1\si^4~,~\ka^D_2\si^2~,~1\r ~,
\nonumber
\\
\fr{G^{L\bar L}}{\lam_L}(M_Z)={\rm Diag}\l \ka^L_1\si^4~,~\ka^L_2\si^2~,~1\r ~.
\la{GMZ}
\end{eqnarray}
The appropriate RG factors are
\begin{eqnarray}
r_t=\exp \left [ -\fr{1}{16\pi^2}\int^{t_3}_{t_Z}\fr{3}{2}\lam_t^2dt\right ]~,
\nonumber
\\
\ka^D_1=\exp \left [ \fr{1}{16\pi^2}\int^{t_3}_{t_1}(\fr{2}{5}g_1^2+8g_3^2)dt\right ]~,
\nonumber
\eeqn
\beqn
\ka^D_2=\exp \left [ \fr{1}{16\pi^2}\int^{t_3}_{t_2}(\fr{2}{5}g_1^2+8g_3^2)dt\right ]~,
\nonumber
\\
\ka^L_1=\exp \left [ \fr{1}{16\pi^2}\int^{t_3}_{t_1}(\fr{9}{10}g_1^2+\fr{9}{2}g_2^2)dt\right ]~,
\nonumber
\\
\ka^L_2=\exp \left [ \fr{1}{16\pi^2}\int^{t_3}_{t_2}(\fr{9}{10}g_1^2+\fr{9}{2}g_2^2)dt\right ]~,
\nonumber
\\
{\rm with}~~~t=\ln \mu ~,~~~~t_z=\ln M_Z~,~~~t_i=\ln \mu_i~,
\nonumber
\\
\mu_1=M_I\si^4~,~~~\mu_2=M_I\si^2~,~~~\mu_3=M_I .
\la{G-RGfactors}
\end{eqnarray}
In  $G^{D\hat{D}}$ and in $G^{L\bar L}$ the renormalization factors are due to gauge interactions.
 The reason is that the heavy states $M_{D_i}$ and $M_{L_i}$ decouple at different scales. Below the decoupling scale
the mass of the corresponding state `freezes' out and does not run, and the leg in the diagram  involving the decoupled state will
not be `dressed' by the loop.

Now we turn to the RG of the $A$ couplings:  At low energies instead of one $A$ matrix we are dealing in general with  six $A$ matrices:
$A\to \l A^{d\hat{d}}, A^{d\hat{D}} , A^{D\hat{d}}, A^{e^-e^+}, A^{e^-E^+}, A^{E^-e^+}\r $. The superscripts indicate
the channels they couple to.
Compared to the $G$-couplings, the $A$ matrices are formed at relatively high scales.These matrices are the only source for generation mixing, but through  loop diagrams involving the scalar states $H_A^6(6,3,1)$ and $H_A^{\bar 6}(\bar 6,\bar 3,1)$, these matrices could  nevertheless induce off diagonal entries into the $G$ matrices. However, if the latter states are already  decoupled
when the $A$ matrices are formed, the corresponding loops will not contribute to renormalization. This is one of the reasons why we take the masses of the
$L_{F'}+\bar L_{\bar F'}$ states equal to $M_6$.  It insures that in the lepton sector non diagonal $A$-couplings
do not  influence the $G$ matrices through loop effects. The $A$-matrices corresponding to the quark channels are formed at the GUT scale keeping the model intact since quarks do not couple with $H_A^6, H_A^{\bar 6}$ .

For the renormalization of the $A$-matrices the relevant energy range is $M_Z-M_I$ where the $G$-matrices are already formed
(above the scale $M_I$ the matrices $A$  do not change their structural form).
Only  the coupling $\lam_t$ is relevant.
Since only the quarks $q_3=(u, d)_3$ are  connected with $\lam_t$, solely the (3,2) elements of the matrices $A^{d\hat{d}}$
and $A^{d\hat{D}}$ will pick up relative RG factors. The spacing of the other entries and of the remaining
$A$ matrices will remain unchanged. Therefore, we have at the scale $M_Z$
\begin{eqnarray}
\fr{A^{d\hat{d}}}{\lam_A^{d\hat{d}}}=\fr{A^{d\hat{D}}}{\lam_A^{d\hat{D}}}={\rm i}\left(
\begin{array}{ccc}
0&\sigma&-\sigma\\
-\sigma&0&1/2\\
\sigma&-\ka_A/2&0
\end{array}\right)~,
\nonumber
\\
\nonumber
{\rm with } \hs{1.7cm} \ka_A=r_t^{-1/3}~\hs{1cm}
\end{eqnarray}
and
\begin{eqnarray}
\fr{A^{D\hat{d}}}{\lam_A^{D\hat{d}}}=\fr{A^{e^-e^+}}{\lam_A^{e^-e^+}}=\fr{A^{e^-E^+}}{\lam_A^{e^-E^+}}=
\fr{A^{E^-e^+}}{\lam_A^{E^-e^+}}=
\nonumber
\\
{\rm i}\left(
\begin{array}{ccc}
0&\sigma&-\sigma\\
-\sigma&0&1/2\\
\sigma&-1/2&0
\end{array}\right)~.~\hs{1cm}
\la{AMZ}
\end{eqnarray}
The numerical values of the RG factors appearing above are
\beqn
r_t= 0.717~,~~~~~~\ka_A= 1.117~,
\non
\\
\non
\ka^D_1= 1.299~,~~~~~~\ka^D_2= 1.127~,
\\
\ka^L_1= 1.137~,~~~~~~\ka^L_2= 1.066 ~.
\la{RGfact-values}
\eeqn
These coefficients are used in the analysis of the charged fermion mass matrices in sect. \ref{sec:3}.

Finally, using the above results,  we are able to fix the values
of the SM singlet VEVs
\beqn
e^3_3=\fr{M_I}{g_2(M_I)}\simeq 2.27\cdot 10^{13}~{\rm GeV}~,
\non
\\
f^3_2=g^3_2=9.5\cdot 10^7~{\rm GeV}~.\hs{0.5cm}
\la{largeVEVs}
\eeqn
On the other hand, from the neutrino sector (see sect. \ref{sec:neutrino}) we have
\beqn
F^{\{2,3\}}=x_g\fr{(e^1_1)^2}{m} \simeq 1.8\cdot 10^{13}~{\rm GeV}~,
\non
\\
F^{\{3,3\}}=x_g\rho F^{\{2,3\}}\simeq 6.6\cdot 10^{11}~{\rm GeV}~,
\non
\\
F_A=x_A \fr{\si^5}{\lam_A}F^{\{2,3\}}\approx -3.3\cdot 10^5~{\rm GeV}~.
\la{FrestYuk}
\eeqn
We recall that by integrating out the $N, \tl{N}$ states, we found for the mass scales in (\ref{FrestYuk})
\beq
F^{\{2,3\}}=\fr{\lam_N\tl{\lam }_N}{M_N}e^3_3\ep^3_2~,~~
F^{\{3,3\}}=-\fr{\lam_N\tl{\lam }_N}{M_N}z(e^3_3)^2 ,
\la{derFscales}
\eeq
(the couplings $\lam_N$, $\tl{\lam }_N$ come from the vertices $\lam_NFH^{\dagger }N$, $\tl{\lam }_NF\tl{H}^{\dagger }\tl{N}$).
The comparison of (\ref{largeVEVs}) with (\ref{FrestYuk})  allows the estimate
\beq
\fr{M_N}{\lam_N\tl{\lam }_N}\simeq 2.9\cdot 10^{13}~{\rm GeV},~
-ze^3_3\hs{-0.08cm}=\hs{-0.08cm}\fr{F^{\{3,3\}}}{F^{\{2,3\}}}\ep^3_2\simeq 8.5\cdot 10^{11}~{\rm GeV}.
\la{MN-tle33-scales}
\eeq
One recognizes again the importance of the scale $M_I = g_2 (M_I)~e^3_3 $  for the heavy fermions and in particular for the 'right handed' neutrinos.

The VEV configuration discussed here is consistent with the charged fermion sector as well as the neutral lepton sector. The following
picture of the  symmetry breaking  of our grand unified model emerges: In a first step  $E_6$ is broken  by an $H(650)$ scalar field.
It reduces $E_6$ to the group $G_{333}=SU(3)_L\tm SU(3)_R\tm SU(3)_C$.
At $M_I=g_2(M_I)e^3_3$ the $G_{333}$ symmetry breaks down to the SM.  The role of $H(650)$ fields is important for the phenomenology. These fields include the $G_{333}$ singlet
 $S_+(1,1,1)$ state. The VEV $\lan S_+\ran =M_{\rm GUT}/g_{\rm GUT}$
gives the breaking $E_6\to G_{333}$ in such a way  that the $L-R$ symmetry ${\cal D}$ is unbroken at high energies. This way a
self consistent gauge coupling unification is guaranteed. The additional SM singlet VEVs are much smaller
than $e^3_3$ . They are important for the fermion sector. The leading role for the electro-weak
symmetry breaking is played by the VEVs  $e_1^1$, $\ep^2_2$, $g^2_3$ and $f^2_3$. The remaining electroweak VEVs ($f^2_2$, $g^2_2$) are small but still
relevant for the light fermion masses and their mixings.

\renewcommand{\theequation}{B.\arabic{equation}}\setcounter{equation}{0}
\section{Neutrino Mass Matrix Renormalization}\label{ApB}

We need to define $m_{\nu }$ at the scale $M_Z$ . To calculate the corresponding renormalization effects it is convenient to consider the combination $({\cal U}^T)^{-1}m_{\nu }{\cal U}^{-1}$ taking
 (\ref{mnu0}),  (\ref{x_g,x_A}) and (\ref{neu-mnu-params}) into account.  Without RG effects this matrix has the form
\beq
\begin{array}{ccc}
 & {\begin{array}{ccc}
 & &
\end{array}}\\ \vspace{1mm}
({\cal U}^T)^{-1}m_{\nu }{\cal U}^{-1}=\hs{-0.2cm}
\begin{array}{c}
 \\  \\
 \end{array}\!\!\!\!\!\hs{-0.2cm} &{\left(\begin{array}{ccc}
\rho &-{\rm i} & {\rm i}
\\
-{\rm i}  &\rho-2x_A  & -{\rm i}\fr{\si }{2}+x_A
 \\
{\rm i}  &-{\rm i}\fr{\si }{2}+x_A &\rho
\end{array}\right)}\cdot m~.
\end{array}  \!\!  ~
\label{UTneuU}
\eeq
In order to compute the RG factors we  recall the origin of each entry in (\ref{UTneuU}). Let's start with the
$\rho m$-entries in the diagonal part. The elements (1,1), (2,2) and (3,3) emerge from dimension five
($d=5$) operators at different scales (at $F^{\{ 2, 3\}}\si^8$, $F^{\{ 2, 3\}}\si^4$ and $F^{\{ 2, 3\}}$, respectively)
by integrating out the appropriate states of $L^3_2, L^3_3$
(this can be seen from Fig. \ref{fig-d5ops}a).
Above the corresponding scale only the Dirac Yukawa
couplings run. After the formation of the effective dimension five operators, i.e. below the corresponding scale,  new RG factors are relevant. They describe the running of couplings of $d=5$ operators which are different from the. running of the square of the Dirac Yukawa couplings. Therefore,
relative (mismatching) RG factors between different entries have to be applied. In order to calculate these RG factors, we need to go
through the entire energy interval considering the appropriate thresholds.
After performing these procedures stepwise from $M_I$
down to $M_Z$  the diagonal $   \rho m $-entries at $M_Z$ can be parametrized as follows
\beq
\rho m(1+r_1)~,~~~~\rho m(1+r_2)~,~~~~{\rm and}~~~\rho m~,
\la{m-MZ-par}
\eeq
where the RG factors $r_{1,2}$ are given by
\beqn
r_1=\exp \left [ \fr{1}{16\pi^2}\int_{t_{11}}^{t_{33}} \l \fr{3}{2}g_2^2+\fr{9}{10}g_1^2\r dt\right ]-1~,
\non
\\
r_2=\exp \left [ \fr{1}{16\pi^2}\int_{t_{22}}^{t_{33}} \l \fr{3}{2}g_2^2+\fr{9}{10}g_1^2\r dt\right ]-1~,
\non
\\
{\rm with}~~~~t_{ii}=\ln \mu_{ii}~,~~~~\mu_{11}=F^{\{ 2, 3\}}\si^8~,\hs{0.7cm}
\non
\\
\mu_{22}=F^{\{ 2, 3\}}\si^4~,~~~~\mu_{33}=F^{\{ 2, 3\}}\hs{0.7cm}
\la{r12-RG-expr}
\eeqn
(when RG effects are ignored these factors are zero and the diagonal $  \rho m$-part of the matrix (\ref{UTneuU}) is a unit matrix
 in flavor space). In the parametrization of (\ref{m-MZ-par}) we did not write any RG factor for the third term $\rho m$
because an overall factor can be absorbed by the parameter $\rho $ (or the scale $m$). Only the relative RG factors given in (\ref{r12-RG-expr}) are needed.

Considering now the remaining entries of $ m_{\nu }$,  it is easy to verify that the (1,2), (1,3) elements together with the
$2 m x_A$ and $m x_A$ entries of (2,2) and (2,3)  are formed at the same
scales after the subsequent integration of the heavy neutral states
(this can be seen from the second and third terms of (\ref{OmMOm-gen}) and
the corresponding diagram in Fig.\hs{0.1cm}\ref{fig-d5ops}.b). Therefore,
no relative RG factors between these entries will appear. The only additional factor occurs for the $-{\rm i}m\fr{\si }{2}$
term in the (2,3)  and (3,2) elements. We can parameterize this term at the scale $M_Z$ by
$-{\rm i}m\fr{\si }{2}r_{23}$. The RG factor $r_{23}$ can be calculated from the equation
\beq
r_{23}=\exp \left [ -\fr{1}{16\pi^2}\int_{t_{11}}^{t_{22}} \l \fr{3}{2}g_2^2+\fr{9}{10}g_1^2\r dt\right ]~.
\la{r23-RG-expr}
\eeq

Taking all RG factors into  account,  the matrix (\ref{UTneuU}) at $M_Z$ takes the form
\beqn
\left.({\cal U}^T)^{-1}m_{\nu }{\cal U}^{-1}\right |_{\mu =M_Z}=\hs{2cm}~
\non
\\
\begin{array}{ccc}
\begin{array}{c}
 \\  \\
 \end{array}\!\!\!\!\!\hs{-0.2cm} &{\left(\begin{array}{ccc}
\hs{-1.5mm}\rho (1+r_1) &-{\rm i} & {\rm i}
\\
\hs{-1.5mm}-{\rm i}  &\rho (1+r_2)-2x_A   & -{\rm i}\fr{\si }{2}r_{23}+x_A
 \\
\hs{-1.5mm}{\rm i} &-{\rm i}\fr{\si }{2}r_{23}+x_A &\rho
\end{array}\hs{-2mm}\right)}\hs{-1.2mm}\cdot \hs{-0.8mm}m~.
\end{array}  \!\!  ~
\label{UTneuU-MZ}
\eeqn
Finally, one obtains the neutrino mass matrix at the scale $\mu =M_Z$ as shown by Eq. (\ref{mnuren}).

\bibliographystyle{unsrt}

\newpage

\onecolumngrid

\begin{figure}[!]
\hspace{0.8cm}
\begin{center}
\leavevmode
\leavevmode
\vspace{2.5cm}
\includegraphics{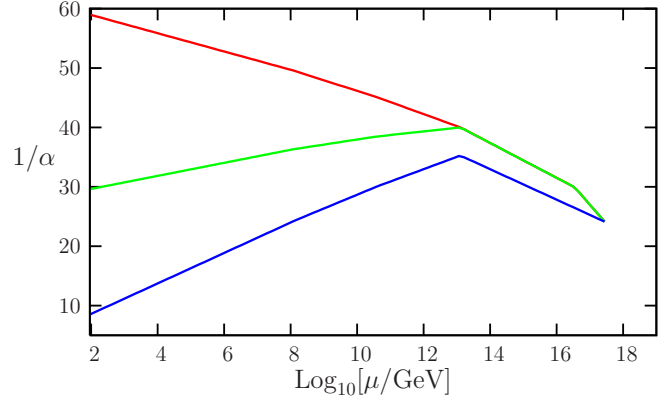}
\end{center}
\vs{3cm}
\caption{'Concorde' -
Unification of gauge couplings.
$M_I\simeq 1.27\cdot 10^{13}$~GeV, $M_6\simeq 3.47\cdot 10^{16}$~GeV,
$M_{\rm GUT}\simeq 2.8\cdot 10^{17}$~GeV and $\al_{\rm GUT}^{-1}\simeq 24.1$.
}
\label{fig-concorde}
\end{figure}

%

%
\vs{4cm}

\begin{center}
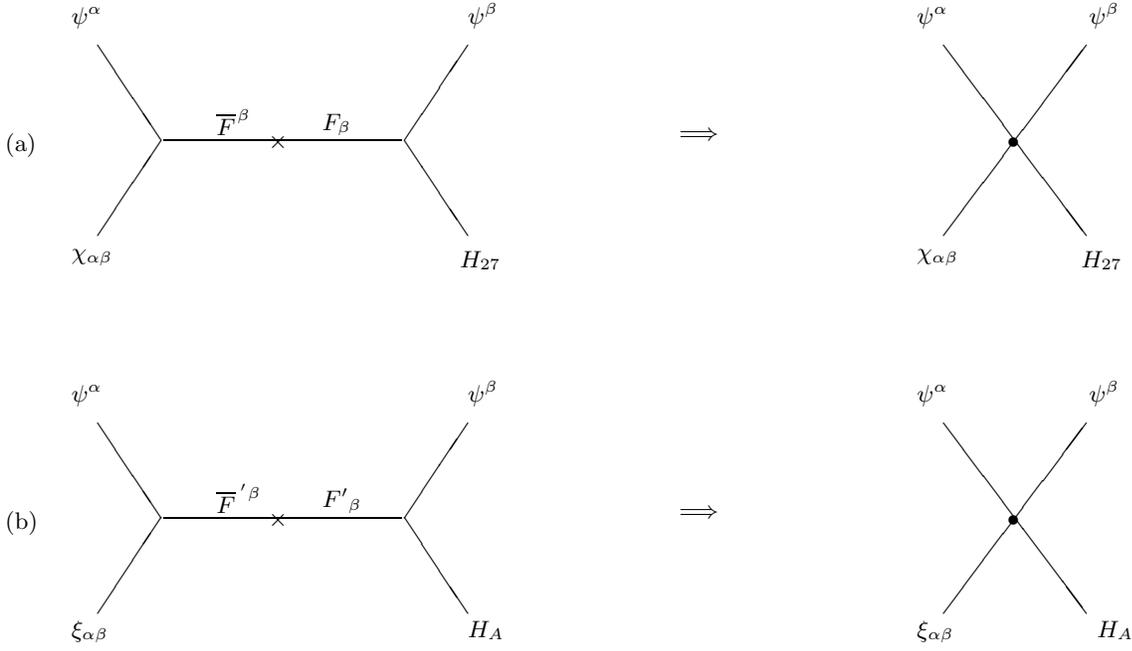
\begin{figure}

\begin{picture}(500,167)(10,-15)

\vs{-8cm}

\put (60,8){\line(2,3){24}}
\put (60,80){\line(2,-3){24}}
\put (85,44){\line(1,0){90}}
\put (176,44){\line(2,3){24}}
\put (176,44){\line(2,-3){24}}

\put (25,40){(a)}
\put (50,-1){$\chi_{\al \bt }$}
\put (50,88){${\psi }^{\al }$}
\put (124.6,41){$\tm $}
\put (200,88){$\psi^{\bt }$}
\put (197,-4){$H_{27}$}
\put (105,46){$\ov{F}^{\hs{0.2mm}\bt }$}
\put (145,48){$F_{\bt }$}

\put (280,44){$\Longrightarrow $}

\put (380,8){\line(3,4){54}}
\put (380,80){\line(3,-4){54}}
\put (404,41){$\bullet $}

\put (370,-1){$\chi_{\al \bt }$}
\put (370,88){${\psi }^{\al }$}
\put (435,88){$\psi^{\bt }$}
\put (432,-4){$H_{27}$}

\end{picture}

\vspace{1.3cm}
\begin{picture}(500,25)(10,115)

\put (60,58){\line(2,3){24}}
\put (60,130){\line(2,-3){24}}
\put (85,94){\line(1,0){90}}
\put (176,94){\line(2,3){24}}
\put (176,94){\line(2,-3){24}}

\put (25,90){(b)}
\put (50,49){$\xi_{\al \bt }$}
\put (50,138){${\psi }^{\al }$}
\put (124.6,91){$\tm $}
\put (200,138){$\psi^{\bt }$}
\put (200,49){$H_A$}
\put (105,96){$\ov{F}^{\hs{0.6mm}'\hs{0.1mm}\bt }$}
\put (145,98){${F'}_{\bt }$}

\put (280,94){$\Longrightarrow $}

\put (380,58){\line(3,4){54}}
\put (380,130){\line(3,-4){54}}
\put (404,91){$\bullet $}

\put (370,49){$\xi_{\al \bt }$}
\put (370,138){${\psi }^{\al }$}
\put (435,138){$\psi^{\bt }$}
\put (438,49){$H_A$}

\end{picture}

\vspace{3cm}
\caption
{Diagrams responsible for operators $\chi_{\al \bt }\psi^{\al }\psi^{\bt }H_{27}$
and $\xi_{\al \bt }\psi^{\al }\psi^{\bt }H_A$.
\label{fig-Fexch}
}

\end{figure}
\end{center}


\begin{center}
\begin{figure}

\vs{-0.2cm}

\begin{picture}(500,80)(10,40)

\put (45,44){\line(-2,3){34}}
\put (45,44){\line(-2,-3){34}}
\put (45,44){\line(1,0){230}}
\put (275,44){\line(2,3){34}}
\put (275,44){\line(2,-3){34}}
\put (115,44){\line(0,-1){50}}
\put (205,44){\line(0,-1){50}}
\put (75,41){$\tm $}
\put (235,41){$\tm $}
\put (155,41){$\tm $}

\put (8,-20){$\chi $}
\put (308,-20){$\chi $}
\put (8,102){${\psi } $}

\put (305,102){$\psi $}
\put (110,-23){$H_{27}^{\dagger }$}
\put (200,-23){$\tl{H}_{27}^{\dagger }$}
\put (60,49){$\ov{F}$}
\put (93,49){$F$}
\put (220,49){$F$}
\put (255,49){$\ov{F}$}
\put (135,49){{\small $N $}}
\put (180,49){{\small $\tl{N}$}}

\put (332,44){$\Longrightarrow $}

\put (380,-5){\line(3,4){70}}
\put (380,88){\line(3,-4){70}}
\put (413,42){\line(3,-2){45}}
\put (370,12){\line(3,2){45}}
\put (412,38){$\bullet $}

\put (375,98){${\psi } $}
\put (445,98){$\psi $}
\put (375,-20){$H_{27}^{\dagger }$}
\put (445,-20){$\tl{H}_{27}^{\dagger }$}
\put (358,5){$\chi $}
\put (462,5){$\chi $}

\end{picture}

\vspace{3cm}
\caption
{Diagram responsible for $\chi^2(H_{27}^{\dagger }{\psi })(\tl{H}_{27}^{\dagger }{\psi })$ operator.
\label{fig-Nexch}
}

\end{figure}
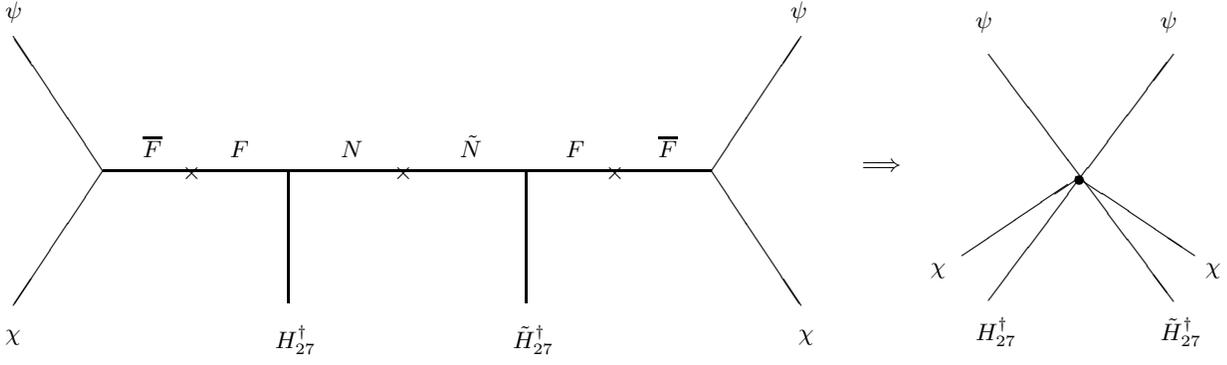
\end{center}


\begin{center}
\begin{figure}

\vs{-1cm}

\begin{picture}(500,35)(10,120)

\put (30,8){\line(2,3){24}}
\put (30,80){\line(2,-3){24}}
\put (55,44){\line(1,0){200}}
\put (256,44){\line(2,3){24}}
\put (256,44){\line(2,-3){24}}

\put (-5,40){(a)}
\put (13,-6){$(L^2_3)^{\al }$}
\put (18,86){$e^1_1$}
\put (105,41){$\tm $}
\put (155,41){$\tm $}
\put (205,41){$\tm $}
\put (278,86){$e^1_1$}
\put (275,-6){$(L^2_3)^{\al }$}
\put (75,52){$(L^3_2)^{\al } $}
\put (126,52){$(L^3_3)^{\al } $}
\put (176,52){$(L^3_3)^{\al } $}
\put (222,52){$(L^3_2)^{\al } $}

\put (302,44){$\Longrightarrow $}

\put (350,8){\line(3,4){54}}
\put (350,80){\line(3,-4){54}}
\put (374,41){$\bullet $}

\put (338,-6){$(L^2_3)^{\al }$}
\put (340,88){$e^1_1$}
\put (405,88){$e_1^1$}
\put (398,-6){$(L^2_3)^{\al }$}

\put (415,44){$\sim \l G\fr{1}{M_1^T}M_2\fr{1}{M_1}G \r_{\al \al} $}

\end{picture}

\vs{4cm}

\begin{picture}(500,55)(10,120)


\put (30,8){\line(2,3){24}}
\put (30,80){\line(2,-3){24}}
\put (159,44){\line(0,1){38}}
\put (55,44){\line(1,0){200}}
\put (256,44){\line(2,3){24}}
\put (256,44){\line(2,-3){24}}

\put (-5,40){(b)}
\put (16,-6){$L^2_3$}
\put (18,86){$e^1_1$}
\put (155,86){$e^1_1$}
\put (105,41){$\tm $}
\put (275,76){$\otimes $}
\put (205,41){$\tm $}
\put (278,89){$g^3_2$}
\put (275,-6){$L^2_3$}
\put (75,52){$L^3_2$}
\put (126,52){$L^3_3 $}
\put (176,52){$L^2_2 $}
\put (225,52){$L^1_1 $}

\put (302,44){$\Longrightarrow $}

\put (350,8){\line(3,4){54}}
\put (350,80){\line(3,-4){54}}
\put (374,41){$\bullet $}

\put (340,-6){$L^2_3$}
\put (340,88){$e^1_1$}
\put (405,88){$e_1^1$}
\put (400,-6){$L^2_3$}

\put (415,44){$\sim g^3_2 G\fr{1}{M_1^T}G\fr{1}{M_0}A^T  $}

\end{picture}

\vspace{6cm}
\caption
{Diagrams responsible for neutrino $d=5$ operators.
\label{fig-d5ops}
}

\end{figure}
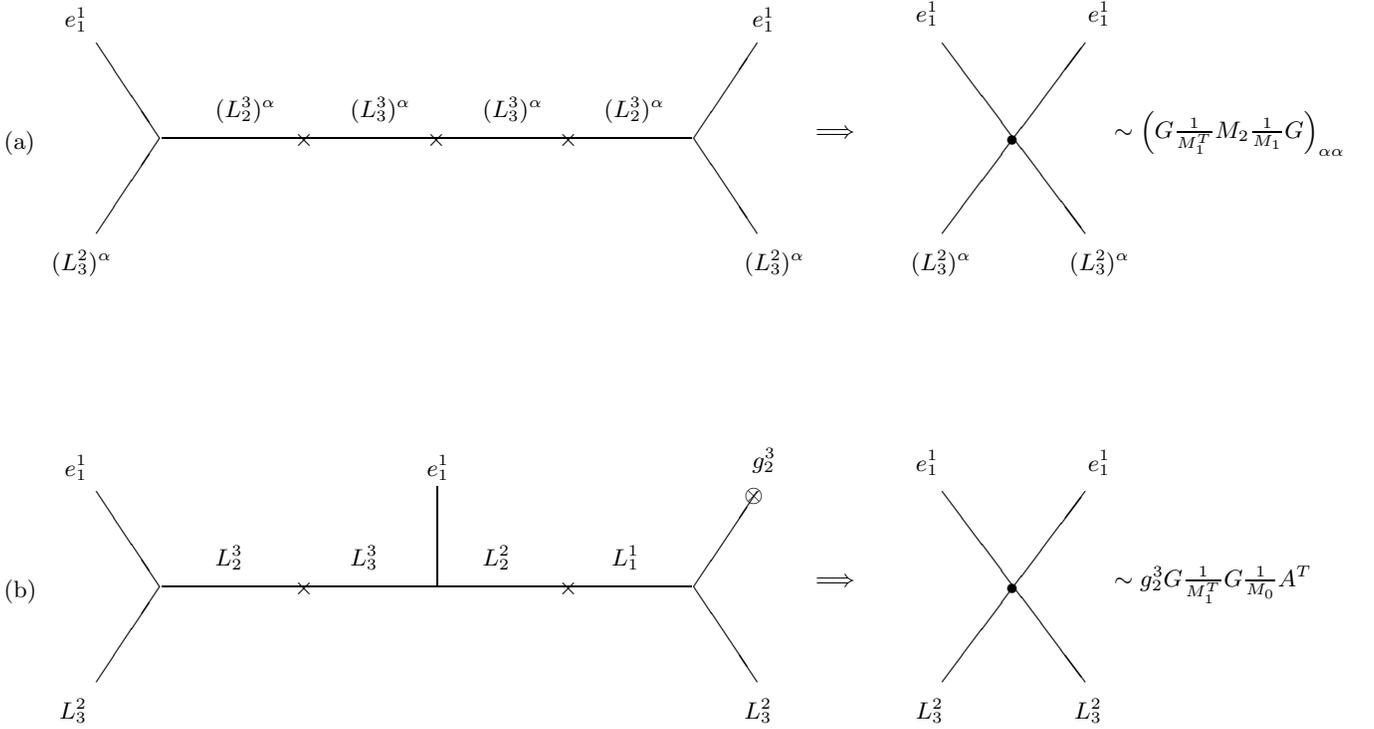
\end{center}

\end{document}